\documentstyle[12pt]{article}

\def\ee{{\rm e}}

\input epsf

\begin{document}
\begin{titlepage}
\rightline{hep-th/yymmddd}
\rightline{EPHOU-95-005}
\rightline{November 1995}
\begin{center}
{\Large {\bf Neutral Multi-Instanton as a Bridge from Weak to Strong
Coupling phase  in Two Dimensional QCD   }}\\
\vspace{2.0cm}
{\large  Tetsuyuki OCHIAI }\\
\vspace{0.3cm}
e-mail : ochiai@particle.phys.hokudai.ac.jp  \\
\vspace{0.3cm}
{\it Department of Physics \\ Hokkaido University \\ Sapporo 060 ,
Japan }
\end{center}
\vspace{4.0cm}

\abstract{
Using a contour integral representation
we analyze the multi-instanton sector in two dimensional
$U(N)$ Yang-Mills theory
 on a sphere  and argue the role of
multi-instanton in the large $N$ phase transition.
In the strong coupling region at the large $N$ , we encounter ``singular
saddle point''. Because of this situation,  ``neutral'' configurations of
the multi-instanton
are dominant in this region.
Based on the ``neutral'' multi-instanton approximation we numerically
 calculate the multi-instanton amplitude ,
the free energies and the Wilson loops for finite $N$ .
We also compare our results with
the large $N$ exact solution
of the free energy and the Wilson loop   and  argue  some problems.
We find the ``neutral'' multi-instanton
contribution bridges the gap between weak and strong coupling phase.
}
\end{titlepage}

\section{Introduction}

In the past years, various attempts to solve low energy QCD have
done. Among them key idea is a kind of string description of QCD because
the string description automatically implies the confinement.
Toward this end , there was remarkable progress for QCD in two dimension
in the last few years.
In particular Gross and Taylor discovered the string descripton of
two dimensional $SU(N)$ or $U(N)$ Yang-Mills theory in terms of $1/N$
expansion\cite{gt}.
It is well known that  the partiotion function ( and the Wilson loop)
of the Yang-Mills theory
 on genus $G$ Riemann surface $\Sigma_G$ can be expressed by
the form of the sum over
irreducible representations  of the gauge group\cite{mr}.
The form  is  called the heat
kernel representation .
They showed that  expanding the heat kernel  by  $1/N$ ,
one gets the sum over the branched covering maps of the Riemann
surface $\Sigma_G$ weighted by
$N^{2-2g}\exp(-\frac{\lambda}{2} A_{{\rm cover }}) $.
Here $g$ and $A_{{\rm cover }}$ are genus and area of the covering space
respectively.
That is, a no-fold string theory with target space  $\Sigma_G$.
But later it was discover by Douglas and Kazakov that on a sphere
this system undergoes
a 3rd order phase transition
 in the large $N$ limit\cite{dk}.
They showed whereas in the strong coupling phase the large $N$ solution
of the free energy
has the string expansion corresponded to the map from sphere to sphere ,
in the weak coupling phase the large $N$ solution is trivial and has no such
expansion.
Thus it is impossible to  interpret QCD as the  string
theory in the weak coupling phase.

Viewed from the strong coupling phase ,the string sum becomes divergent
at the phase transition point due to the entropy of the branch points\cite{t}.
So the entropy of the branch points physically causes the phase transition.
On the other hand ,viewed from the weak coupling phase
instanton induces the phase transition\cite{mp,gm1}.
We recall Witten's work on two dimensional QCD\cite{w}.
He showed that the Yang-Mills partition function on Riemann surface
can be expressed by a sum over the instanton.
The weight associated with the instanton is corectly determined from the heat
kernel representation\cite{mp}.
Using this formula Gross-Matytsin  showed  the following results\cite{gm1}.
The 0 instanton sector gives the weak coupling result fot the free energy.
 In the weak coupling region the 1 instanton amplitude of charge $\pm 1$
is strongly suppresed in the large $N$ limit but at the phase transition point
the  damping factor disappears .

In addition one can see using their results
whereas the damping factor is disappearded at the phase transition point,
the 1 instanton amplitude of charge $\pm 1$ still has  order $N^{-1/2}$
in the strong coupling region.  And it's contribution to the free energy is
order $N^{-5/2}$ and thus negligible in the large $N$ limit.
Hence the 1-instanton effect is insufficient to explain the large $N$ phase
transition and it is important to investigate the effect of
the multi-instanton
in the large $N$ limit.
That is our motivation for analyzing the multi-instanton sector.

Since in the strong coupling phase Yang
-Mills theory has the string description, by studying the phase transition
 we can see how the nonperturbative
effect of the multi-instanton construct the string picture.
That is instructicve for real four dimensional QCD.
In four dimensional QCD it is unlikely that  there is such phase
transition\cite{gm1}.
But perturbation can not lead to the string picture, it is important to
know how the nonperturbative effect construct the string picture.
By considering the case of two dimension in detail, there might give some
insight for four dimensional  QCD.
We should keep the above thing in mind when we  study the large $N$
phase transition in two dimensional QCD.

In two dimension  the Yang-Mills theory has  no transverse gluon
and  it is in some sense topological.
So coupling to various matter fields is very important.
But such systems are obtained by   considering
 random walk for the Wilson loop of the Yang-Mills theory\cite{s},
our method  might become another
approach to the above systems. \\
\vskip10pt

In the preceeding paper the auther gave a   scenario of the large $N$ phase
transition which explains the $O(1)$ contribution to the free energy
in the strong coupling region using the multi-instanton\cite{o}.
In this paper  following the preceeding paper,  we continue our analysis on
 the multi-instanton contribution to the free enegy and the Wilson
loop and find out that there is a  sort of neutrality
in the strong coupling phase at the large $N$.
That is, ``neutral'' configurations of the
multi-instanton are dominated for both the free energy and the Wilson loop
in this region.
Based on the ``neutral'' multi-instanton scheme, we also calculate the
free energy and the Wilson loop for gauge group $U(3),U(4)$and $U(5)$.
And we compare our results with the large $N$ exact solution of the free
energy\cite{dk} and the Wilson loop\cite{bdk} and find that the neutarl
multi-instanton bridges the gap between the weak and the strong coupling
phase in two dimensional QCD on sphere.

 The content of this paper is as follows.
 In section 2   the formula of the multi-instanton amplitude and the partition
function are obtained in terms of  contour integral.
 In section 3  the formula of  the Wilson loop average are also obtained
in terms of  contour integral.
 In section 4 nature of the 1-instanton and the dilute gas approximation
are considered.
 In section 5 the structure of the multi-instanton amplitude
in the large $N$ limit is analyzed in detail.
 In section 6 numerical calculation of
the multi-instanton amplitude , the free energy and the Wilson loop are
performed.
Also  our results are comared to the large $N$ exact solution .
 In conclusion we sumary the results.

\section{ partition function }

The partition function of two dimensional $U(N)$ Yang-Mills theory
has so called the heat kernel representation.
The representation was first considered by A.Migdal on disk
in the context of the
real space renormalization group and later generalized to
arbitraly 2-manifold by B.Rusakov\cite{mr}.
On a sphere with area $A$ , the partition function becomes
\begin{eqnarray}
    Z(A)&=& \int {\cal D}A_{\mu} \exp(-\frac{N}{4\lambda}\int
                       d^{2} x \sqrt{g}{\rm tr}F_{\mu \nu}F^{\mu\nu})
                        \nonumber \\
   &=&\sum_{R} ({\rm dim} R)^{2} \exp (-\frac{\lambda A}{2N} C_{2}(R)) .
\end{eqnarray}
Here
 $R$ is irreducible representation of the gauge group $U(N)$,dim$R$ is the
dimension of representation $R$ and $C_2(R)$ is
the value of the  second Casimir operator of rep $R$.
In terms of the highest weight components of  $R$ , ${n_1\geq n_2
\geq \ldots  \geq n_N}$,
dim$R$ and $C_2(R)$ are given by
\begin{eqnarray}
{\rm dim}R &=& \prod_{1 \leq i \leq j \leq N} (1-\frac{n_i-n_j}{i-j}) , \\
C_2(R) &=& \sum_{i=1}^N n_i (n_i +N+1-2i) .
\end{eqnarray}
If we define $l_i$ as $n_i-i +\frac{N+1}{2}$ , the partition function becomes
\footnote{Hereafter we absorb $\lambda$ into A .}
\begin{equation}
Z(A)={\rm const}\quad \ee^{ \frac{A}{24} (N^2 -1)}
     \sum_{l_1 >\cdots>l_N} \prod_{1 \leq i \leq j \leq N}
        \Delta^2 (l) \exp (-\frac{A}{2N} \sum_{i=1}^N l_i^2 ).
\end{equation}
We remark that the condition that $l_i$'s satisfy the relation
${l_1> l_2> \ldots  > l_N}$
is irrevant because the configuration such as $l_i=l_j$ does not contribute to
the Van der Monde determinant  $\Delta$  and $Z(A)$
has the  index permutation symmetry.
Then the sum becomes free sum over ${\bf Z}^N $ for odd $N$
or sum over $({\bf Z}+\frac{1}{2})^N $  for even $N$ and we can rewrite
it by the Poisson resumation formula. This is a sort of duality
transformation.
We get the following expression
\footnote{A similar expression was obtained by
M. Caselle et al \cite{cdmp}. They showed that the phase
transition is due to the winding modes of ``fermion on circle''.}
\cite{mp}:
\begin{eqnarray}
      Z(A)&=& \ee^{\frac{A}{24}(N^2-1)}({N\over A})^{N^2}
          \sum_{\{m_i\} \in {\bf Z}^N} \epsilon^{\sum_{i=1}^N m_i }
          \ee^{-\frac{2 \pi^2 N}{A} \sum _{i=1}^{N } m_{i}^{2}} w(\{m\}) ,\\
      w(\{m\})&\equiv &\int_{-\infty}^{\infty} \prod_{i=1}^{N} dy_{i}
                          \prod_{i<j}^{N} (y_{ij}^{2} -4\pi^{2} m_{ij}^{2} )
                          \ee^{- \frac{N}{2A} \sum_{i=1}^{N} y_{i}^{2} }
                           \nonumber \\
         &=&\int_{-\infty}^{\infty} \prod_{i=1}^{N} dy_{i}
                  \Delta (y_{i}+2\pi m_{i})\Delta (y_{i}-2\pi m_{i})
                  \ee^{- \frac{N}{2A} \sum_{i=1}^{N} y_{i}^{2} }
\end{eqnarray}
where $y_{ij}=y_i-y_j$, $m_{ij}=m_i-m_j$, $\epsilon =1$ for odd $N$ and
$\epsilon =-1$ for even $N$.
In \cite{mp,gm1,w} it was showed that
$\{m\}$  corresponde to all Euclidean classical solutions
(which we call instanton) up to gauge transformation .
So the instanton is dual to the highest weight component.
The instanton   have nonperturbative effect with respect
to both $1/N$ and $\lambda$. Hereafter we call $m_{i}$
 instanton charge and call number of nonzero $m_{i}$'s  instanton number.
In the following we analyze the phase transition using the duality
and see the instanton induced the phase transition.
That is reminiscent of the
duality transformation of classical XY model and the  Kosterliz-Thouless phase
transition \cite{kt}.
In this  case  viewed from the low temperature the
vortex condensation causes the phase
transition.

The multi-instanton amplitude $ w(\{m\}) $ looks like
the partition function of the
Gaussian  Hermite   matrix model but there is a
deformation in the Van der Monde determinant $\Delta$.
In this section we rewrite $w(\{m\})$  using the method of
ortho-polynomial (in this case Hermite polynomial) and  obtain
a new  contour integral representation.
 The new representation makes clear the role of the multi-instanton in
the large $N$ phase transition.

  First using the property of the Van der Monde determinant
 we obtain  :
\begin{equation}
w(\{m\})
= \sum_{\mu \in S_{N}}  {\rm sgn} \mu \sum_{\sigma \in
              S_{N}}
              \prod_{i=1}^{N}\int_{-\infty}^{\infty} dy_{i}
              P_{\sigma (i)} (y_{i} +2\pi m_{i} )
              P_{\mu \circ \sigma (i)} (y_{i} -2\pi m_{i} )
              \ee^{ - \frac{N}{2A} y_{i}^{2}  } ,
\end{equation}
where $S_N$ is the permutation group on $N$ object and
\begin{equation}
P_n (x)=\frac{1}{2^n (\frac{N}{2A})^{\frac{n}{2}}}H_n(\sqrt{\frac{N}{2A}}x)
 =x^n +( {\rm lower} \quad {\rm power })
\end{equation}
is the ortho-polynomial under the Gaussian weight ;
\begin{equation}
\int_{-\infty}^{\infty} dy P_{n}(y)
                P_{m}(y)  \ee^{- \frac{N}{2A} y^2} =h_{n}\delta_{nm}
                =\sqrt{2\pi}(\frac{A}{N})^{n+\frac{1}{2}} n! \delta_{nm} .
\end{equation}
  In the following let us assume number of non-zero $m_i$'s is $k$ i.e.
number  of instanton  is $k$.
Using the permutation symmetry
nonzero $m_i$'s  are driven to $m_{i}$'s from $i=1$ to $i=k$.
Then we get,
\begin{eqnarray}
  & &      w(m_1,\ldots,m_k,0,\ldots ,0) \nonumber  \\
  &=& \sum_{\mu \in S_{N}} {\rm sgn} \mu \sum_{\sigma \in S_{N}}
              \prod_{i=k+1}^{N} h_{\sigma ( i)}
               \delta_{\sigma ( i), \mu \circ \sigma (i) }
              \prod_{i=1}^{k} \int_{-\infty}^{\infty} dy_{i}
                P_{\sigma (i)} (y_{i} +2\pi m_{i} )
                P_{\mu \circ \sigma (i)} (y_{i} -2\pi m_{i} )
               \ee^{- \frac{N}{2A}  y_{i}^{2} } \nonumber \\
  &=& (N-k)! \sum_{\mu \in S_{k}} {\rm sgn} \mu
                  \sum_{a_{1} \neq \cdots \neq a_{k}}
                  h_{0} \cdots \check{ h_{a_{1}}}  \cdots
                  \check{h_{a_{k}}}\cdots h_{N-1}    \nonumber  \\
  & & \qquad \times
\prod_{i=1}^{k} \int_{-\infty}^{\infty} dy_{i}
                  P_{a_{i}} (y_{i}+2\pi m_{i} )
                  P_{\mu(a_{i})} (y_{i} -2\pi m_{i} )
                  \ee^{- \frac{N}{2A}  y_{i}^{2}  } .
\end{eqnarray}
Here $\mu \in S_{k}$ is the element of the permutation group acting on the
set $\{ a_1,\ldots ,a_k \}$.
Using the  Taylor series expansion of
the Hermite polynomial, we obtain for the above  integral,
\begin{eqnarray}
& & \prod_{i=1}^{k} \int_{-\infty}^{\infty} dy_{i}
                  P_{a_{i}} (y_{i}+2\pi m_{i} )
                  P_{\mu(a_{i})} (y_{i} -2\pi m_{i} )
                  \ee^{- \frac{N}{2A}  y_{i}^{2} }
                   \times (h_{a_{1}} \cdots h_{a_{k}})^{-1} \nonumber \\
&=& \prod_{l=1}^{k} \sum_{i_{l}=0}^{{\rm min} (a_{l},\mu (a_{l}))}
                 \frac{\,_{\mu (a_{l})} C_{i_{l}}}{(a_{l}-i_{l})!}
                (\frac{2\pi m N}{A})^{a_{l}-i_{l}}
                (-2 \pi m)^{\mu (a_{l})-i_{l}} .
\end{eqnarray}
The each series in the above equation can be represented as
a contour integral by the following transformation formula.
\begin{equation}
\sum_{i=0}^{{\rm min}(a,b)}
                 \frac{\,_{b} C_{i}}{(a-i)!}
                (\alpha)^{a-i}  (\beta)^{b-i}
      = \oint \frac{ dt }{2\pi i} \ee^{\alpha \beta t}
                 \frac{1}{t} (\frac{1}{\beta t})^{a}
                 (\beta (t+1) )^{b} ,
\end{equation}
where the contour of $t$ encircles only the origin  counterclockwise. Using
this formula, we obtain the  following contour integral representation
 for the $k$-instanton amplitude.
\begin{eqnarray}
& &    w(m_1,\ldots,m_k,0,\ldots ,0) \nonumber \\
&=& \frac{(N-k)!}{N!}Z_G(\frac{N}{A})
                  \sum_{a_{1} \neq \cdots \neq a_{k}}^{N-1}
                   \sum_{\mu \in S_{k}} {\rm sgn} \mu
                   \oint \frac{ dt_{1} }{2\pi i} \cdots
                   \oint \frac{ dt_{k} }{2\pi i}
                   \frac{1}{t_{1} \cdots t_{k}}  \nonumber \\
& & \qquad\times\ee^{-\frac{ 4 \pi^2 N}{A} (m_{1}^2  t_{1}+\cdots+
                    m_{k}^2 t_{k} ) }
                    (\frac{m_{\mu (1)}}{m_{1}}
                    \frac{1+t_{\mu (1)}}{t_{1}})^{a_{1}}
                    \cdots (\frac{m_{\mu (k)}}{m_{k}}
                    \frac{1+t_{\mu (k)}}{t_{k}})^{a_{k}} .
\end{eqnarray}
Here $Z_G (\beta ) \equiv \int \prod dx \Delta^2 (x)
\exp(-\frac{\beta}{2}\sum x^2 )$ is the partition function of the Gaussian
Hermite matrix model.
We remark the configurations such as $a_1=a_2 $ do not affect the above
equation. Hence we can replace $ \sum_{a_{1} \neq \cdots \neq a_{k}}^{N-1}$
to independent sum $\sum_{a_1}^{N-1}\cdots \sum_{a_k}^{N-1}$.
 In that expression there
are many pole free terms.
After eliminating them, the amplitude  becomes
\begin{eqnarray}
&=& \frac{(N-k)!}{N!}Z_G(\frac{N}{A})
                   \oint \frac{ dt_{1} }{2\pi i} \cdots
                   \oint \frac{ dt_{k} }{2\pi i}
                   \ee^{-\frac{ 4 \pi^2 N }{A}(m_{1}^2  t_{1}+\cdots+
                    m_{k}^2 t_{k} ) }
                  (1+\frac{1}{t_{1}})^{N} \cdots (1+\frac{1}{t_{k}})^{N}
                   \nonumber \\
& & \qquad \times \sum_{\mu \in S_{k}} {\rm sgn} \mu
                   \frac{1}{\frac{m_{\mu (1)}}{m_{1}} (1+t_{\mu (1)})-t_{1}}
                   \cdots
                   \frac{1}{\frac{m_{\mu (k)}}{m_{k}} (1+t_{\mu (k)})-t_{k}} .
\end{eqnarray}
The last sum is just determinant of matrix $M$ which has $(ij)$ element
\begin{equation}
 M_{ij} \equiv  \frac{1}{\frac{m_{j}}{m_{i}} (1+t_{j})-t_{i}} .
\end{equation}
It's diagonal element is 1 and this determinant has all information about the
interaction between instantons.

For the sake of numerical calculation we further rewrite the $k$-instanton
amplitude
 in terms of  another contour integral.
For this purpose, we classify elements of the permutation group $S_{k}$ into
the conjugacy classes. By symmetry argument, the contribution from different
elements belonging to same conjugacy class is same. Then we obtain,
\begin{eqnarray}
 & & \sum_{m_{1},\ldots ,m_{k} \neq 0} \epsilon^{ m_1+\cdots +m_k }
     {\rm e}^{-\frac{ 2 \pi^2 N }{A}(m_{1}^2 +\cdots+ m_{k}^2) }
     w(m_1,\ldots ,m_k,0,\ldots ,0) \nonumber \\
 &=&  \frac{(N-k)!}{N!}Z_G(\frac{N}{A}) \sum_{m_{1},\ldots ,m_{k} \neq 0}
      \epsilon^{ m_1+\cdots +m_k }
      \oint \frac{ dt_{1} }{2\pi i} \cdots \oint \frac{ dt_{k} }{2\pi i}
      {\rm e}^{-N(\widetilde{\Phi}_{m_{1}}(t_{1})+\cdots+
                  \widetilde{\Phi}_{m_{k}}(t_{k})) }
        \nonumber \\
      & & \qquad\qquad \times \sum_{{\rm conj. class}} {\rm sgn}[\sigma ]
         T[\sigma ]
       M_{1\sigma (1)} M_{2\sigma (2)} \cdots M_{k\sigma (k)} ,
\end{eqnarray}
where
\begin{equation}
\widetilde{\Phi}_{m}(t) \equiv \Phi_{m}(t)+\frac{2 \pi^2 m^2 }{A}
      \equiv    \frac{ 4 \pi^2 m^2}{A}(t+\frac{1}{2})-\log(1+\frac{1}{t}).
\end{equation}
We assume $\sigma $ has cycle structure $[1^{\sigma_{1}}\cdots
k^{\sigma_{k}}]  (\sigma_{1} +\cdots +k\sigma_{k} =k)$ and put $T[\sigma ]$
as number of elements in the conjugacy class which  $\sigma $ belongs to,
\begin{eqnarray}
     T[\sigma ] &=& \frac{k!}{1^{\sigma_{1}} \sigma_{1} ! \cdots
                               k^{\sigma_{k}} \sigma_{k} ! } , \\
     {\rm sgn} \sigma &=& (-)^{\sigma_{2}+\sigma_{4}+ \cdots
                               +\sigma_{2[\frac{k}{2}]}} , \\
     \sum_{{\rm conj  class}} &=& \sum_{\sigma_{1},\ldots ,\sigma_{k}=0
                                                   \atop
                                         \sigma_{1} +\cdots +k\sigma_{k}
=k}  .
\end{eqnarray}
We use (12345)(67) type element as the representative element of the
conjugacy class.
 Because the multiple integral factorize according to the cycle structure
of $\sigma$,  the multi-instanton amplitude is  exponentiated with
a constraint $( \sigma_{1} +\cdots +k\sigma_{k} =k) $ which is expressed by
a contour integral of $z$. We get the following result:
\begin{eqnarray}
& & \sum_{m_{1},\ldots ,m_{k} \neq 0} \epsilon^{ m_1+\cdots +m_k }
     {\rm e}^{-\frac{ 2 \pi^2 N }{A}(m_{1}^2 +\cdots+ m_{k}^2) }
     w(m_1,\ldots ,m_k,0,\ldots ,0) \nonumber \\
&=& ({}_N C_k )^{-1} Z_G(\frac{N}{A}) \oint \frac{ dz }{2\pi i}
     \frac{1}{z^{k+1}}
     \exp(\sum_{j=1}^{N} \frac{(-)^{j-1}}{j} z^{j}
     \alpha_{j}(A) ) ,
\end{eqnarray}
where the contour of $z$ encircles only the origin counterclockwise and
\begin{equation}
\alpha_{j}(A)\equiv  \sum_{m_{1},\ldots,m_{j}\neq 0}
      \epsilon^{ m_1+\cdots +m_j }
      \oint \frac{ dt_{1} }{2\pi i} \cdots \oint \frac{ dt_{j} }{2\pi i}
      {\rm e}^{-N(\widetilde{\Phi}_{m_{1}}(t_{1})+\cdots+
                  \widetilde{\Phi}_{m_{j}}(t_{j})) }
        M_{12} M_{23} \cdots M_{j1}
\end{equation}
is the ``connected'' amplitude of the $j$ instantons.
In the eq(21) we  extrapolate the range of summation inside
the exponential  from $\sum_{j=1}^{k}$ to $\sum_{j=1}^{N}$ ,but this
does not affect the equation due to the $z$ integral.

 Using the above equation we finally obtain for the partition function
\begin{eqnarray}
Z(A)&=&\ee^{\frac{A}{24}(N^2-1)}(\frac{N}{A})^{N^2} \sum_{k=0}^N {}_N C_k
     \sum_{m_{1},\ldots ,m_{k} \neq 0} \epsilon^{\sum_{i=1}^k m_i}
     {\rm e}^{-\frac{ 2 \pi^2 N }{A}\sum_{i=1}^k m_i^2 }
     w(m_1,..,m_k,0,..,0) \nonumber \\
 &=& Z_{{\rm weak}}(A) \oint \frac{ dz }{2\pi i}
     \frac{1}{z^{N+1}}
     \frac{1}{1-z} \exp(\sum_{j=1}^{N} \frac{(-)^{j-1}}{j} z^{j}
     \alpha_{j}(A) ) ,
\end{eqnarray}
where
\begin{equation}
Z_{{\rm weak}}(A) \equiv
        \ee^{\frac{A}{24}(N^2-1)}Z_{G}(A/N)=
       {\rm const} \, \ee^{N^2 (\frac{A}{24}-
       \frac{1}{2}\log A ) - \frac{A}{24} }.
\end{equation}
Here we use the the dual property of the Hermite Gaussian matrix model as
$Z_G (\beta)=\beta^{-N^2}Z_G (1/\beta )$. Since $Z_{{\rm weak}}(A)$ is the
0-instanton (i.e. all $m_i=0$ )contribution to the partition function ,
$Z_{{\rm weak}}(A)$  is just the equation (4) before the duality
transformation in which the discrete sum
is replaced by the continious integral.
That is the peculiar property of the Poisson resumation formula.
Douglas-Kazaov showed that   $Z_{{\rm weak}}(A) $  is in fact the partition
function  in the weak coupling phase at the large $N$.
Thus the multi-instanton effect is essential in the strong coupling region
at the  large $N$.

\section{ Wilson loop }

In this section, following the preceeding method ,
 we give the contour integral representation of the multi-instanton
sector in the Wilson loop.
The heat kernel represenatation gives the following expression for the
  Wilson loop on a sphere  \cite{mr} :
\begin{eqnarray}
& & W_{n}(A_1,A_2)
         \equiv \langle \frac{1}{N} {\rm tr}(P \ee^{i \oint A})^n
         \rangle \nonumber \\
&=& \frac{1}{Z(A)}
    \sum_{R,S}{\rm dim} R{\rm dim} S\ee^{-\frac{A_1 C_2 (R)+A_2 C_2 (S)}{2N}}
    \frac{1}{N} \int dU \chi_R (U) \chi_S (U^{\dag}) {\rm tr} U^n .
\end{eqnarray}
Here the loop devide the sphere with area $A$ into two  disks with
area  $A_1,A_2$.
The above holonomy integral gives the linear combination of the
Clebsh-Gordan coefficient. For example of $n=1$ case ,
The sum over representation $R,S$ becomes that over $R$ multiplied by the
sum over all posible  attachment of one box to the Young tableau of $R$.
Hence in the same way as the partition function ,
the sum becomes free sum over ${\bf Z}^N$ for odd $N$ or
$({\bf Z}+\frac{1}{2})^N$ for even $N$.
By the Poisson duality  we get the following result of the Wilson loop average
\cite{gm2}.
\begin{eqnarray}
& & W_{n}(A_1,A_2) Z(A) \nonumber \\
&=&\ee^{\frac{A}{24}(N^2-1)}(\frac{N}{A})^{N^2}
\frac{1}{N}\sum_{k=1}^{N} \sum_{\{m\} \in {\bf Z}^N }
\epsilon^{\sum_{i=1}^{N} m_i}
\ee^{-\frac{2\pi^2 N}{A}\sum_{i=1}^{N} m_{i}^{2}-2\pi i n m_{k}
         \frac{A_2}{A} -\frac{n^2 A_1 A_2}{2NA}} \nonumber \\
& &\times \int_{-\infty}^{\infty} \prod_{i=1}^{N} dy_{i}
                  \Delta (y_{i}+2\pi m_{i}+\frac{inA_2}{N}\delta_{ik})
                  \Delta (y_{i}-2\pi m_{i}+\frac{inA_1}{N}\delta_{ik})
                  \ee^{-\frac{N}{2A} \sum_{i=1}^{N} y_{i}^{2}} .
\end{eqnarray}
where $\frac{1}{N}\sum_{k=1}^{N}\exp(-2\pi i n m_{k}\frac{A_2}{A})$
is the classical value of
the Wilson loop with the multi-instanton configuration.
{}From the above equation, the Wilson loop average apparently has $A_1
\leftrightarrow A_2 $ exchange symmetry. We also find that
calculating the Wilson loop is like
calculating the  partition function with additional imaginary charged
instanton.
Hence in the same way as the partition function,  we can rewrite this
in terms of a contour integral.
In the $l$-instanton sector ( i.e. number of non-zero m's is $l$ ) ,
we devide the place   where the imaginary chage put on .
One is $k$ with nonzero $m_k$,  another is $k$ with zero $m_k$.
Hence the above equation becomes
\begin{eqnarray}
&=& \ee^{\frac{A}{24}(N^2-1)}(\frac{N}{A})^{N^2}
         \ee^{-\frac{n^2 A_1 A_2}{2AN}}
         \frac{1}{N} \sum_{l=1}^{N} {}_{N} C_{l}\sum_{m_1,\cdots ,m_l \neq 0}
         \epsilon^{m_1+\cdots +m_l}
         \ee^{-\frac{2 \pi^2 N}{A} (m_1^2+\cdots +m_l^2 )} \nonumber \\
& & \times [ l \ee^{-2\pi i n m_{1} \frac{A_2}{A}}
             \int_{-\infty}^{\infty} \prod_{i=1}^{N} dy_{i}
             \ee^{-\frac{N}{2A} \sum_{i=1}^{N} y_{i}^{2}}    \nonumber \\
& &  \quad  \times  \Delta (y_1+2\pi m_1+inA_2 /N,y_2+2\pi m_2 ,\cdots ,
                           y_l+2\pi m_l ,y_{l+1},\cdots ,y_N) \nonumber \\
& &  \quad \times   \Delta (y_1-2\pi m_1+inA_1 /N,y_2-2\pi m_2 ,\cdots ,
                           y_l-2\pi m_l ,y_{l+1},\cdots ,y_N) \nonumber \\
& &  +(N-l) \int_{-\infty}^{\infty} \prod_{i=1}^{N} dy_{i}
          \ee^{-\frac{N}{2A} \sum_{i=1}^{N} y_{i}^{2}}      \nonumber \\
& &  \quad  \times    \Delta (y_1+2\pi m_1,\cdots ,y_l+2\pi m_l ,
                           y_{l+1}+inA_2 /N ,y_{l+2},\cdots ,y_N) \nonumber \\
& &  \quad  \times \Delta (y_1-2\pi m_1,\cdots , y_l-2\pi m_l ,
                           y_{l+1}+inA_1 /N ,y_{l+2},\cdots ,y_N) ].
\end{eqnarray}
The first integral in the big parentheses, using the preceeding method, has the
following contour integral representation :
\begin{eqnarray}
& & \frac{(N-l)!}{N!} Z_G (\frac{N}{A})
    \oint \frac{ dt_{1} }{2\pi i} \cdots \oint \frac{ dt_{l} }{2\pi i}
    \ee^{-\frac{ 4 \pi^2 N }{A}(m_{1}^2  t_{1}+\cdots+m_{l}^2 t_{l} ) }
    (1+\frac{1}{t_1})^{N} \cdots (1+\frac{1}{t_l})^{N}  \nonumber \\
& & \qquad\qquad \times \ee^{ 2\pi i n m_1 \frac{A_1-A_2}{A} t_1
                              -\frac{n^2}{N}\frac{A_1 A_2}{A} t_1}
                        {\rm det}^{(l)}B .
\end{eqnarray}
The second integral also has the contour integral representation as
\begin{eqnarray}
& & \frac{(N-l-1)!}{N!} Z_G (\frac{N}{A})
   \oint \frac{ dt_{1} }{2\pi i} \cdots \oint \frac{ dt_{l+1} }{2\pi i}
   \ee^{-\frac{ 4 \pi^2 N }{A}(m_{1}^2  t_{1}+\cdots+m_{l}^2 t_{l})  }
    (1+\frac{1}{t_1})^{N} \cdots (1+\frac{1}{t_l})^{N}  \nonumber \\
& &  \qquad\qquad \times
     \ee^{-\frac{n^2}{N}\frac{A_1 A_2}{A}t_{l+1}}(1+\frac{1}{t_{l+1}})^{N}
                        {\rm det}^{(l+1)}C  ,
\end{eqnarray}
where
\begin{eqnarray}
B_{ij}&=& \frac{1}{\frac{b_j}{b_i}(1+t_{j})-t_{i}} \quad  ,\quad
b_i =-2\pi m_i +\frac{in A_1}{N} \delta_{i1} ,   \\
C_{ij}&=& \frac{1}{\frac{c_j}{c_i}(1+t_{j})-t_{i}}  \quad  ,\quad
c_i = -2\pi m_i (1-\delta_{i,l+1})+\frac{in A_1}{N} \delta_{i,l+1}.
\end{eqnarray}
For the sake of numerical calculation , we further rewrite it to a
exponential form.
The first term of the $l$-instanton contribution to the Wilson loop
 has the following form:
\begin{equation}
\sum_{m_1,\cdots ,m_l \neq 0} \epsilon^{m_1+\cdots +m_l}
\oint \prod_{i=1}^l \frac{ dt_{i} }{2\pi i} {\rm
e}^{-N\widetilde{\Phi}_{m_{i}}(t_{i})}
f(m_1,t_1) {\rm det }^{(l)} B .
\end{equation}
In the determinant we assume the cycle including the index 1 to have length
$k$
and explicitly to be $(1,a_1,\ldots ,a_{k-1})$ .Thus the determinant will be
\begin{equation}
{\rm det}^{(l)} B = \sum_{k=1}^l \sum_{a_1 \neq \cdots a_{k-1} \neq 1}
               B_{1 a_1}\cdots B_{a_{k-1} 1} {\rm det }^{(l-k)}B ,
\end{equation}
here $ {\rm det }^{(l-k)}B$ is the determinant not including the index
$ 1,a_1 ,\ldots a_{k-1}$.
Under the index rename invariant measure ,
 we can rename $a_1 ,\ldots ,a_{k-1}$
 to be $ 2, \ldots ,k $. Thus we get
\begin{equation}
{\rm det}^{(l)}B \to
\sum_{k=1}^l \frac{(l-1)!}{(l-k)!}(-1)^{k-1}
 B_{12} B_{23} \cdots B_{k1} {\rm det}^{(l-k)} B
\end{equation}
Hence eq.(32) becomes
\begin{eqnarray}
& & \sum_{m_1,\cdots ,m_l \neq 0} \epsilon^{m_1+\cdots +m_l}
    \oint \prod_{i=1}^l \frac{ dt_{i} }{2\pi i}
    {\rm e}^{-N\widetilde{\Phi}_{m_{i}}(t_{i})}
    f(m_1,t_1) {\rm det }^{(l)}B  \nonumber \\
&=& l! \sum_{k=0}^{l-1} \frac{1}{k!}\beta_{l-k}(A_1,A_2)
    \sum_{m_1,\cdots ,m_k \neq 0} \epsilon^{m_1+\cdots +m_k}
    \oint \prod_{i=1}^k \frac{ dt_{i} }{2\pi i}
    {\rm e}^{-N\widetilde{\Phi}_{m_{i}}(t_{i})}{\rm det}^{(k)}M .
\end{eqnarray}
where
\begin{eqnarray}
\beta_{k}(A_1,A_2)&\equiv& (-1)^{k-1}\sum_{m_1,\cdots ,m_k \neq 0}
                \epsilon^{m_1+\cdots +m_k}
                \oint \prod_{i=1}^k \frac{ dt_{i} }{2\pi i}
                {\rm e}^{-N\widetilde{\Phi}_{m_{i}}(t_{i})}
                 B_{12} B_{23} \cdots B_{k1} \nonumber \\
& & \quad \times  \ee^{-2\pi i n m_1 \frac{A_2}{A}
                                +2\pi i n m_1 \frac{A_1-A_2}{A} t_1
                                -\frac{n^2}{N}\frac{A_1 A_2}{A} t_1}
\end{eqnarray}
is the type 1 connected amplitude of $k$-instanton

 In the same way , under the index rename invariant measure ,we get
\begin{equation}
{\rm det}^{(l+1)}C \to
\sum_{k=1}^{l+1} \frac{l!}{(l-k+1)!}(-1)^{k-1}
 C_{12} C_{23} \cdots C_{l+1 1} {\rm det}^{(l+1-k)} C .
\end{equation}
Hence the second term of the $l$-instanton contribution to the Wilson
loop becomes
\begin{eqnarray}
& &  \sum_{m_1,\cdots ,m_l \neq 0} \epsilon^{m_1+\cdots +m_l}
    \oint \prod_{i=1}^l \frac{ dt_{i} }{2\pi i}
    {\rm e}^{-N\widetilde{\Phi}_{m_{i}}(t_{i})}
    \oint \frac{ dt_{l+1}}{2\pi i}\ee^{-\frac{n^2 A_1 A_2}{NA}t_{l+1}}
   (1+\frac{1}{t_{l+1}})^{N} {\rm det }^{(l+1)}C  \nonumber \\
&=& l! \sum_{k=0}^{l} \frac{1}{k!}\gamma_{l-k}(A_1,A_2)
    \sum_{m_1,\cdots ,m_k \neq 0}\epsilon^{m_1+\cdots +m_k}
    \oint \prod_{i=1}^k \frac{ dt_{i} }{2\pi i}
    {\rm e}^{-N\widetilde{\Phi}_{m_{i}}(t_{i})}{\rm det}^{(k)}M .
\end{eqnarray}
where
\begin{eqnarray}
\gamma_{k}(A_1,A_2)&\equiv &(-1)^k \sum_{m_1,\cdots ,m_k \neq 0}
            \epsilon^{m_1+\cdots +m_k}
            \oint \prod_{i=1}^{k+1} \frac{ dt_{i} }{2\pi i}
            \prod_{i=1}^k {\rm e}^{-N\widetilde{\Phi}_{m_{i}}(t_{i})}
                    \nonumber \\
& & \qquad \times  \ee^{-\frac{n^2 A_1 A_2}{NA}t_{k+1}}
            (1+\frac{1}{t_{k+1}})^{N}
            C_{12} C_{23} \cdots C_{k+1 1} .
\end{eqnarray}
is the type 2 connected amplitude of k-instanton.
Using the fact that $k$-instanton amplitude can be rewritten as eq (21),
we finaly obtain the following result:
\begin{eqnarray}
& &  W_{n}(A_1,A_2) Z(A) /
       Z_{{\rm weak}}(A)  \ee^{-\frac{n^2 A_1 A_2}{2AN}}   \nonumber \\
&=& \frac{1}{N} \sum_{l=0}^N \oint \frac{dz}{2\pi i}
     ( \sum_{k=0}^{l-1}\frac{\beta_{l-k}(A_1,A_2)}{z^{k+1}}+
       \sum_{k=0}^{l} \frac{\gamma_{l-k}(A_1,A_2)}{z^{k+1}} )
      \exp (\sum_{j=1}^{N} \frac{(-)^{j-1}}{j} z^{j} \alpha_{j}(A) ) .
\end{eqnarray}
This is the base of our numerical calculation of the Wilson loop average.

For example , in the 0-instanton sector  only
$\gamma_0 $ contributes to the Wilson loop. We get
\begin{equation}
  W_{n}(A_1,A_2) Z(A) /
     Z_{{\rm weak}}(A)  \ee^{ -\frac{n^2 A_1 A_2}{2AN}}
     |_{0-{\rm inst}}
= \frac{1}{N} \gamma_0 =
    \frac{1}{N} \oint \frac{dt}{2\pi i} \ee^{-\frac{n^2 A_1 A_2}{NA}t}
     (1+\frac{1}{t})^{N} .
\end{equation}
By rescaling  $t \to Nt $, we obtain the known result of the Wilson loop
in the weak coupling
phase at the large $N$  \cite{bdk} :
\begin{equation}
  W_{n}(A_1,A_2)|_{0-{\rm inst}} =
     \frac{1}{n}\sqrt{\frac{A}{A_1 A_2}}J_{1}(2n
     \sqrt{\frac{A_1 A_2}{A}}) +O(1/N) .
\end{equation}
So the multi-instanton effect is essential for the Wilson loop in the strong
coupling phase.
Apparently if we extrapolate the above equation
 to $A \to \infty$ \footnote{The phase
transition  occures before $A \to \infty $.} with fixed $A_1$ ,
 we cannot obtain the simple area law on infinite disk but obtain a
oscilating behavior.

\section{Effects of 1-instanton}

In this section  we consider the contribution of one instanton  to the
partition function\cite{gm1} and the Wilson loop. We see that the
1-instanton effect is insufficient to explain the large $N$  phase
transition.
We use the word ``1-instanton'' to mean that only $\alpha_1 ,\beta_1,
\gamma_0 ,\gamma_1$ are  included in the
partition function and the Wilson loop.
So this is a kind of dilute gas  approximation in our system.

The dilute gas approximation leads to the following result for the partition
function.
\begin{equation}
\frac{Z(A)}{Z_{{\rm weak}}(A)}\to \oint \frac{ dz }{2\pi i}
     \frac{1}{z^{N+1}}
     \frac{1}{1-z} \exp( z\alpha_1 (A) ) \nonumber \\
= \sum_{k=1}^{N} \frac{(\alpha_1 (A))^k}{k!}.
\end{equation}
We should note since maximum number of the instanton is $N$ , the partition
function in this approximation is not completely exponentiated as
$\exp(\alpha_1)$.
In the large $N$ limit, we can apply the saddle point method to
$\alpha_1$. The saddle point equation $\Phi_m^{\prime}(t)=0$
 has two solutions
\begin{equation}
t=t_{\pm}\equiv \frac{-1\pm \sqrt{1-\frac{A}{\pi^2 m^2}}}{2}.
\end{equation}
We see quite different behavior according to $A {> \atop <} \pi^2 m^2$.

For the case of $A < \pi^2 m^2$,
both saddle points  exist on negative real axis and satisfy
\begin{equation}
\Phi_{m}(t_{-})< \Phi_{m}(t_{+}) .
\end{equation}
Neverthless only $t_{+}$ saddle point is selected
since $t_{-}$ steepest decent line is not consistent with the original
contour. Hence we obtain in the large $N$ limit ,
\begin{eqnarray}
\oint\frac{dt}{2 \pi i}\ee^{-N\widetilde{\Phi}_m(t)}
& \simeq &
(-)\frac{1}{\sqrt{2\pi N |\Phi_{m}^{(2)}(t_{+})|}}
                     {\rm e}^{-N\widetilde{\Phi}_{m}(t_{+})} \nonumber \\
&=&\frac{(-)^{N-1}}{\sqrt{2\pi N \frac{16\pi^4m^4}{A^2}
\sqrt{1-\frac{A}{\pi^2m^2}}}}
{\rm e}^{-\frac{2\pi^2 m^2N}{A}
                   \gamma(\frac{A}{\pi^2m^2})} .
\end{eqnarray}
where  $\gamma (x)$ is defined as \cite{gm1}
\begin{equation}
\gamma (x)\equiv
\sqrt{1-x}-\frac{x}{2}\log\frac{1+\sqrt{1-x}}{1-\sqrt{1-x}} ,
\end{equation}
Since  $\gamma (x)$ is positive real for $x<1$  , in this region
1-instanton amplitude is strongly suppresed with order
$\frac{1}{\sqrt{N}}\ee^{-N}$ .

For the case of  $A > \pi^2 m^2$,
 both saddle points must be selected because $t_{\pm}$ steepst
descent lines are consisitent with original contour
and satisfy
\begin{equation}
{\rm Re}[\Phi_{m}(t_{+})]={\rm Re}[\Phi_{m}(t_{-})] .
\end{equation}
Therefore, we get
\begin{equation}
\oint\frac{dt}{2 \pi i }\ee^{-N\widetilde{\Phi}_m(t)} \simeq
\frac{(-)^N}{\sqrt{2\pi N \frac{16\pi^4
m^4}{A^2} \sqrt{\frac{A}{\pi^2 m^2}-1}}}
  ({\rm e}^{i\frac{3}{4}\pi
                 -\frac{2\pi^2 m^2N}{A}
                   \gamma(\frac{A}{\pi^2m^2})}
  +{\rm e}^{-i\frac{3}{4}\pi
                 +\frac{2\pi^2 m^2N}{A}
                   \gamma(\frac{A}{\pi^2m^2})}) .
\end{equation}
Since $\gamma (x) $ are pure imaginary in $x>1$, the above equation
has order $ N^{-1/2}$ in the large $N$ limit.

{}From the above  observation  the following truncation about
the instanton charge
is justified.
If $A$ satisfies the relation $k^2 \pi^2 < A < (k+1)^2 \pi^2$ ($k$ :
positive integer ), we
truncate the instanton charge as $-k \leq m\leq k$.
That is , According to $A$ , the instanton charges which are
included in the large $N$ limit are changed stepwise as figure1.
\vskip10pt
\centerline{\epsfysize=2in\epsffile{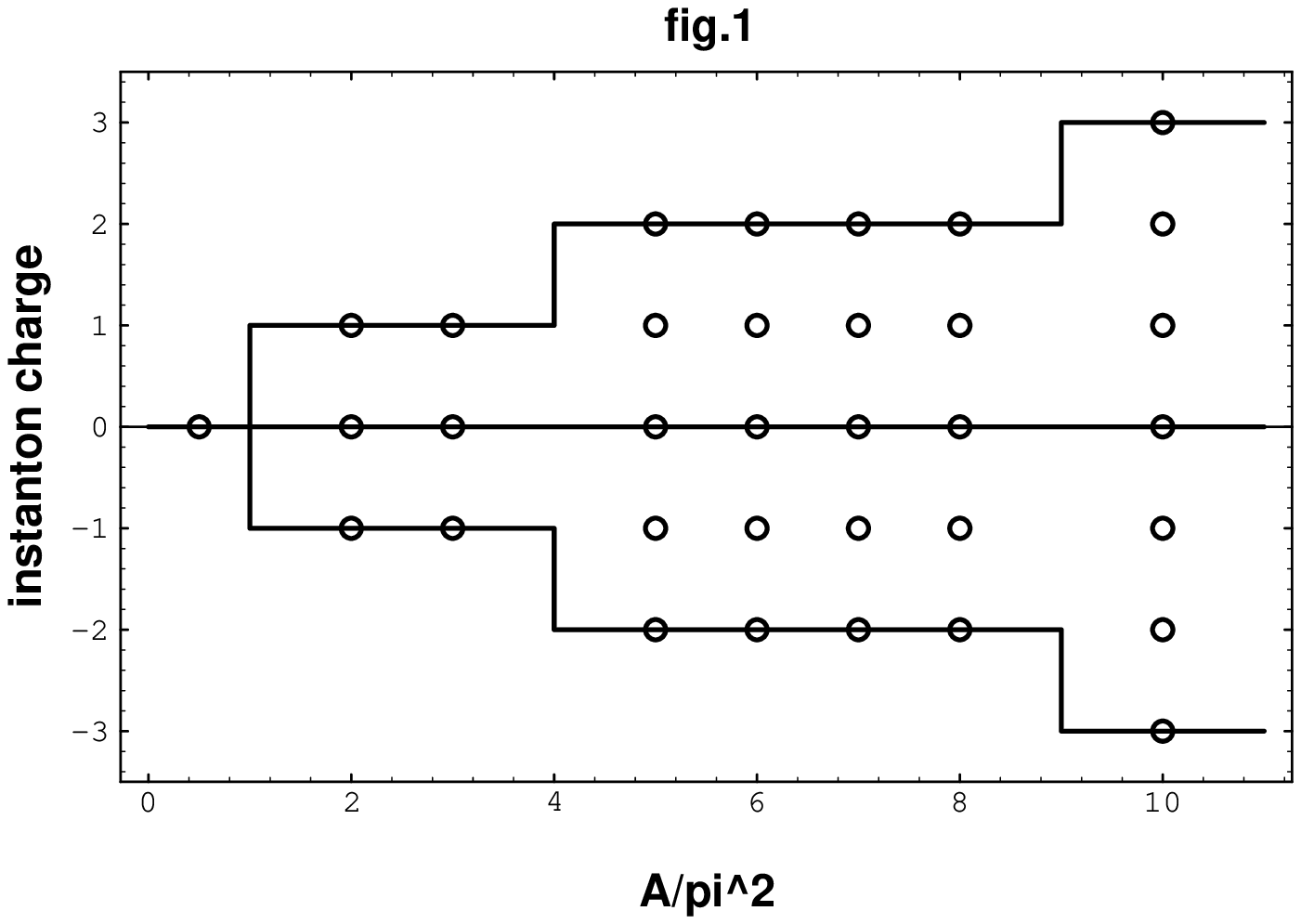}}\vskip4pt
\hangindent\parindent{{\small {\bf Figure1}:
The instanton charges which must be included in the large $N$ limit (indicated
by circle )
are changed  stepwise according to $A$. }}
\vskip10pt
This truncation is basically justified in the multi-instanton amplitude
because $\alpha_k$ become factorized as $(\alpha_1)^k$ in the interaction
free approximation. But we will see in the next section a peculiar
dynamics in the multi-instanton interaction  make further truncation
justified in the large $N$ limit.

Since $\alpha_1$ has order $N^{-1/2}$ in $A\ge \pi^2$,we may replace
eq (43) to
$\exp(\alpha_1)$.But this gives order $N^{-5/2}$ result for the free energy
$\widetilde{F}\equiv \frac{1}{N^2}\log[Z/Z_{\rm weak}]$.
To see explicitly that the dilute gas approximation is insufficient
we show in figure 2 the $N=5$ result of the free energy with the dilute gas
approximation  and the large $N$ Douglas-Kazakov solution which is
described later .
\vskip10pt
\centerline{\epsfysize=2in\epsffile{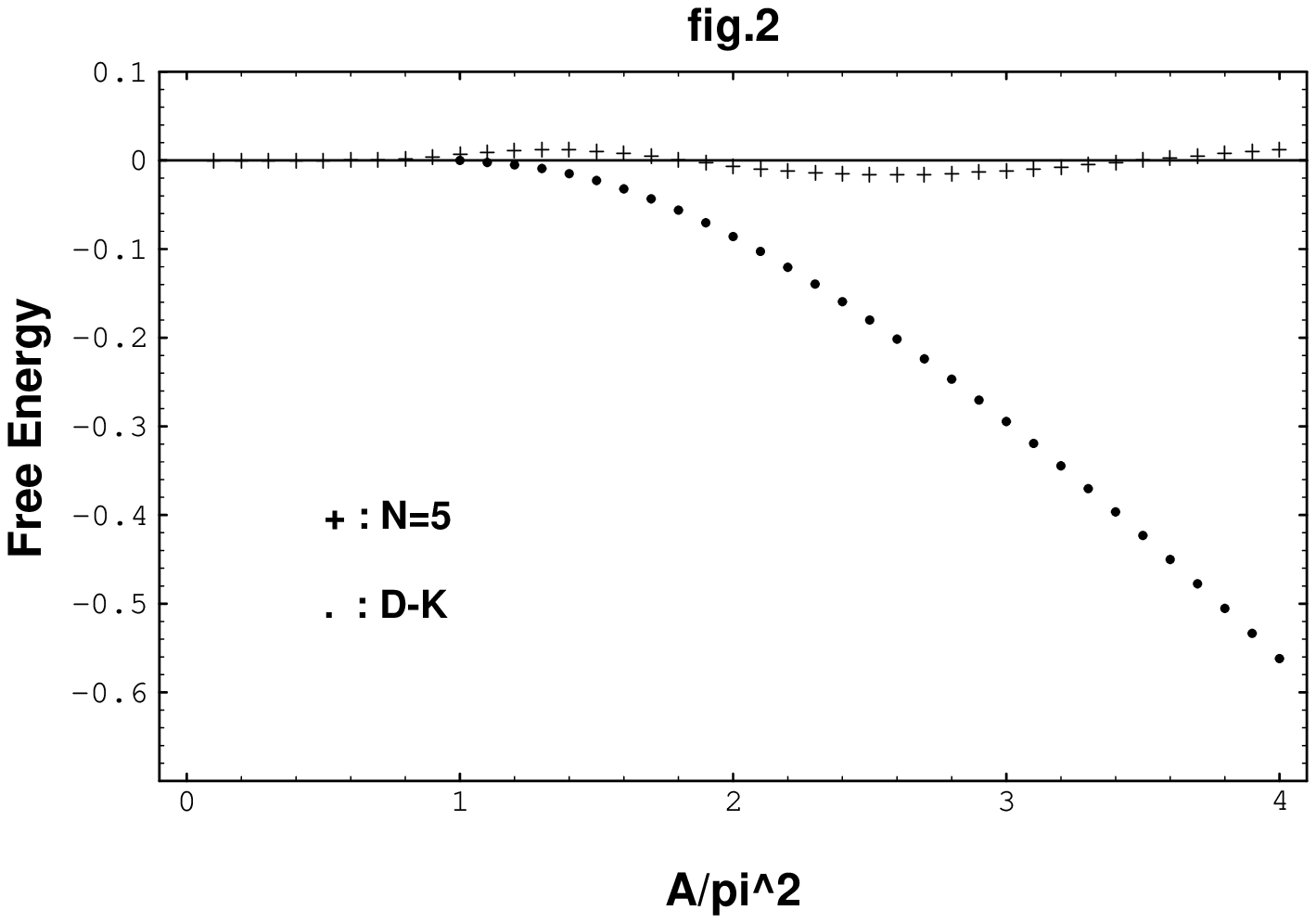}}\vskip4pt
\hangindent\parindent{ {\small {\bf Figure2}:
The free energy of $N=5$ in the dilute gas approximation and the large $N$
D-K solution. }}
\vskip10pt

Next we consider the dilute gas approximation of the Wilson loop.
This lead to
\begin{equation}
W_n(A_1,A_2)\to [ \frac{1}{N}\gamma_0 (A_1,A_2) +
     \frac{1}{N} (\beta_1(A_1,A_2)
    +\gamma_1(A_1,A_2)) \frac{ \sum_{k=1}^{N-1} \frac{(\alpha_1 (A))^k}{k!}}
 {\sum_{k=1}^{N} \frac{(\alpha_1 (A))^k}{k!}} ]
  \ee^{ -\frac{n^2 A_1 A_2}{2AN}}
\end{equation}
for the Wilson loop.
Thus we consider $\beta_1$ and $\gamma_1$.
In the same way as $\alpha_1$ ,the large $N$ saddle point method leads to
$\beta_1 ,\gamma_1 ={\it O}(\frac{1}{\sqrt{N}}\ee^{-N})$ for $ A < \pi^2 $
and
$\beta_1 ,\gamma_1 ={\it O}(\frac{1}{\sqrt{N}})$ for $ A > \pi^2 $.
Here we use the rescaling $t_2 \to Nt_2$ for $\gamma_1$.
Thus in the large $N$ limit the dilute gas approximation lead to the value
in the weak coupling phase. To see explicitly the insufficiency of the dilute
gas approximation we show in figure 3 the $N=5$ results of the Wilson loop
average
\begin{equation}
\widetilde{W}_1(A_1,A_2)\equiv
W_1(A_1,A_2)- \frac{1}{N}\gamma_0 (A_1,A_2)\ee^{ -\frac{ A_1 A_2}{2AN}}
\end{equation}
in the approximation and compare the results with the large
$N$  Boulatov,Daul-Kazakov  (BDK) solution
\begin{equation}
{\widetilde{W}_1}^{\rm BDK} (A_1,A_2)\equiv
{W_1}^{\rm BDK} (A_1,A_2)-\sqrt{\frac{A}{A_1 A_2}}J_{1}(2
     \sqrt{\frac{A_1 A_2}{A}}) .
\end{equation}
\vskip10pt
\centerline{\epsfysize=2in\epsffile{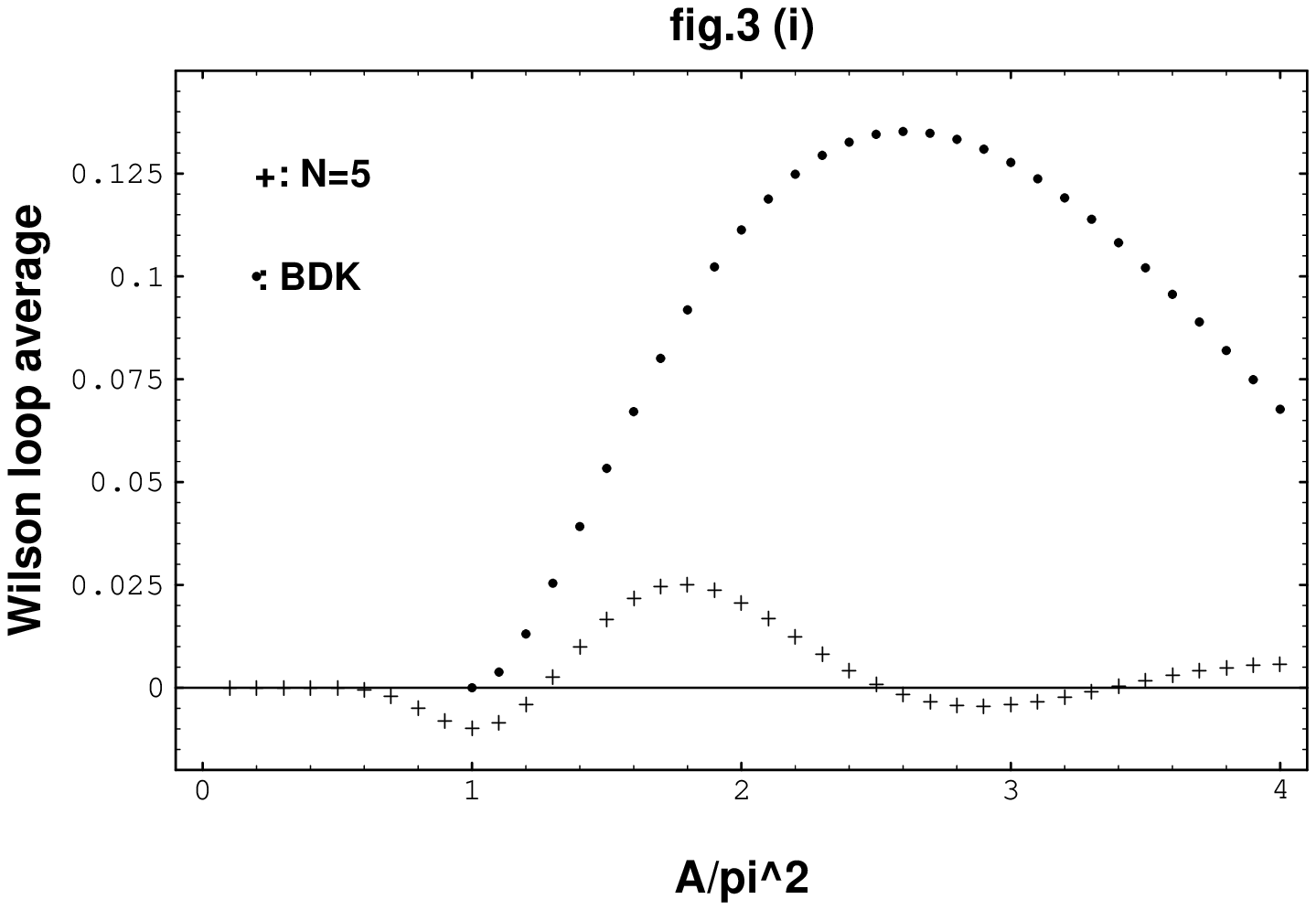} \hskip2pt
            \epsfysize=2in\epsffile{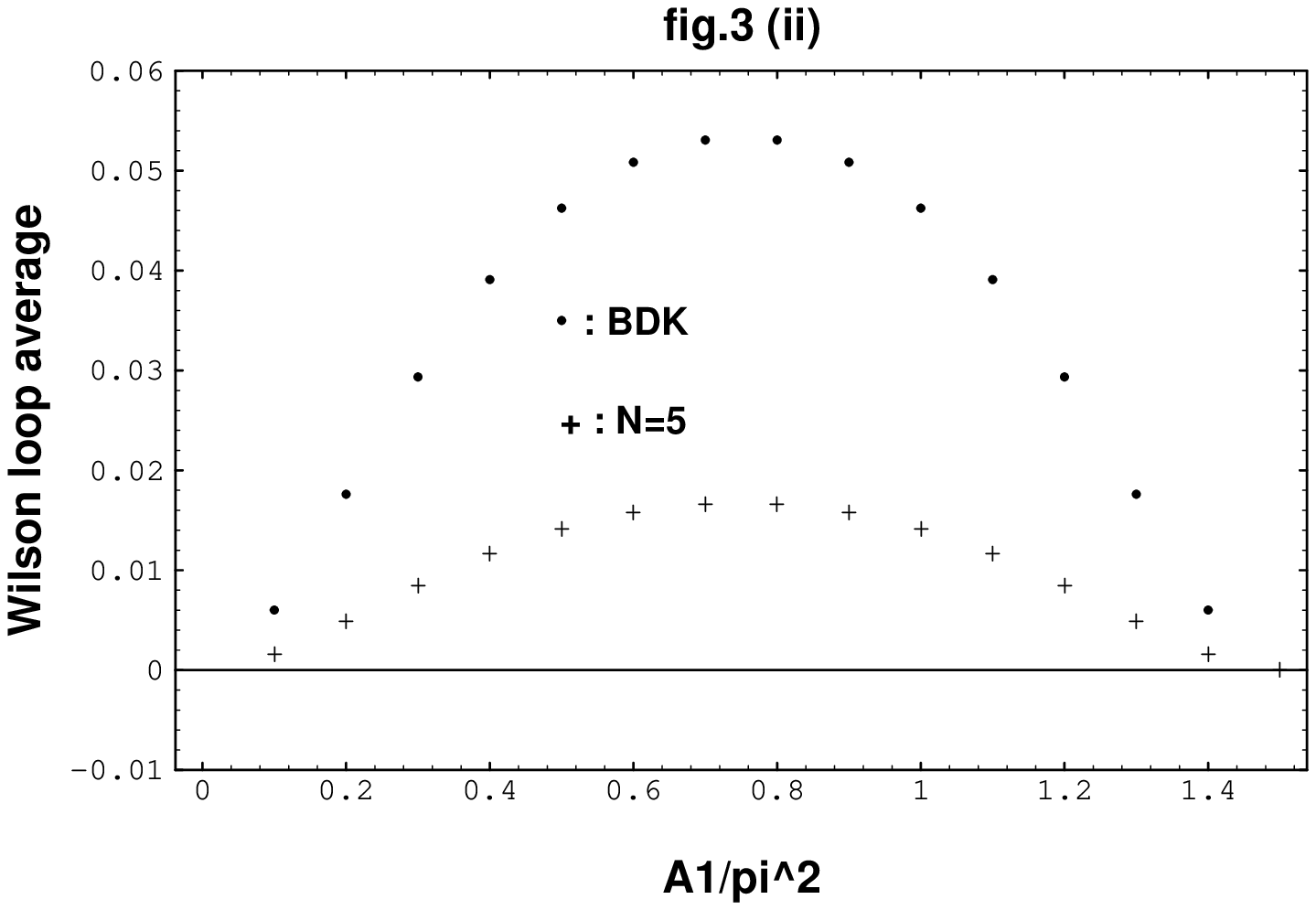}}
\vskip4pt
\hangindent\parindent{ {\small {\bf Figure3}:
The Wilson loop average of $N=5$ in the dilute gas approximation and
the large $N$ BDK solution  with (i) $A_1=A/2$ (ii) $A=1.5\pi^2$ }}
\vskip10pt

\section{Large $N$ neutrality }

In the preceeding sections we gave the contour integral representation for the
multi-instanton sector of the partition function and the Wilson loop. The
representation has a simple gauge group parameter ($N$) dependence. That is ,
$\alpha , \beta, \gamma$  have the form such as :
\begin{equation}
\oint \prod dt \ee^{-N f(t)} g(t) .
\end{equation}
where $g(t)$ depends not on $N$ for $\alpha_k$ and  weakly depends on $N$ for
$\beta_k , \gamma_k$.
Then ,If we consider  the large $N$ limit, the integrals for
 $\alpha,\beta,\gamma$ are dominated by the saddle points given by the
solution of $ \partial f(t)=0$.
Fortunately, it is no need for  solving the complicated saddle point equations
corresponded to the  multi-dimensional contour integral in our case.
The saddle point equations are
decoupled for each variables and the solutions of the equations are sets of
each one dimensional solution which  was
found by Gross and Matytsin \cite{gm1}.
Then if we neglect the interaction term
$M_{12} M_{23} \cdots M_{j1}$ ( for example $\alpha_j$ 's case ) ,
$\alpha_j$ becomes factorized such as
$(\alpha_1)^j$ and the behavior change from
exponential damping to oscilating  at $A=\pi^2 m^2 $ observed in
the preceeding section still hold for the multi-instanton amplitude.
But there is a remarkable feature in  the interaction term of
the multi-instanton sector.

 We point out that in the region of $A>\pi^2$  special charge
configurations of the multi-instanton exist.
For this purpose let us first
consider the following toy contour integral:
\begin{equation}
\oint \frac{dt}{2\pi i} \ee^{-N\phi (t)} \frac{1}{(t-t_{c})^{n}} ,
\end{equation}
where $t_{c}$ is the saddle point of $\phi (t)$.
We can approximate the above equation in the large $N$ limit
 to a principal value
integral along the steepest
descent line and a contour integral along small semi circle around the  saddle
point. As the result we  obtain finite value with order $N^{\frac{n-1}{2}}
\exp(-N\phi (t_c))$.
We call this type of problem ``singular saddle point problem ''.
Apparentry, the degree of singularity in  $\frac{1}{(t-t_{c})^{n}}$
at the saddle point determines the power of $N$.

In our case such problem occures in a complicated manner.
Let's consider the singularity of $\alpha_k$.
The maximal singularity  occures when $t$'s are the solution of
$M_{ii+1}^{-1}=0$. That is
\begin{equation}
\left( \begin{array}{ccccc}
-1             &\frac{m_2}{m_1}& 0             & \cdots & 0     \\
0              &  -1           &\frac{m_3}{m_2}&    0   & 0     \\
\vdots         &               &               &        &\vdots \\
\frac{m_1}{m_k}&   0           & \cdots        &   0    &  -1
\end{array} \right)
\left( \begin{array}{cc}
t_1    & \\
t_2    & \\
\vdots & \\
t_k    &
\end{array} \right)  =
\left( \begin{array}{cc}
-\frac{m_2}{m_1}    & \\
-\frac{m_3}{m_2}    & \\
\vdots      & \\
-\frac{m_1}{m_k}    &
\end{array} \right) .
\end{equation}
The consistency of the above equation requires $m_1+\cdots +m_k=0$ i.e.
neutral configuration.
Under this condition the equation has a one parameter solution
given by
\begin{equation}
\left( \begin{array}{cc}
t_1    & \\
t_2    & \\
\vdots & \\
t_{k-1}& \\
t_k    &
\end{array} \right)  =
\left( \begin{array}{cc}
\frac{m_k}{m_1} (t+1)+\frac{m_{k-1}+\cdots +m_2}{m_1}   & \\
\frac{m_k}{m_2} (t+1)+\frac{m_{k-1}+\cdots +m_3}{m_2}   & \\
\vdots                                                  & \\
\frac{m_k}{m_{k-1}} (t+1)                               & \\
t                                                       &
\end{array} \right) .
\end{equation}

For the case of even $k$ if we consider the following
configurations
\begin{equation}
(m_{1},m_{2},m_{3},m_{4},\ldots ,m_{k})=(m ,-m,m,-m,\ldots ,-m)
\end{equation}
which we call ``neutral'',
the above solution simply becomes
\begin{equation}
t_{{\rm even}} =t \quad , \quad t_{{\rm odd}}=-(t+1).
\end{equation}
On the other hand ,under the above charge configuration ,
the large $N$ saddle points for any $t_{i}$ are same and given by :
\begin{equation}
t_{\pm}=\frac{-1\pm \sqrt{1-\frac{A}{\pi^2 m^2}}}{2} \quad
, \quad t_{+}+t_{-}+1=0 .
\end{equation}
When $A>m^2 \pi^2$, both saddle points must be selected for each variable.
 Hence  the saddle
points include the following combinations:
\begin{equation}
t_{{\rm even}}=t_{\pm} \quad , \quad t_{{\rm odd}}=t_{\mp}=-(t_{\pm}+1).
\end{equation}
In this case  the above saddle points ride on the line of maximal
singularity and
thus we encounter the singular saddle point problem.
We can also show the ``neutral'' configuration is  the only case
in which the saddle point completely
rides on the line of the maximal singularity.
Hence the neutral  configuration  are
dominant in the large $N$ limit of $\alpha_k$ .
Compared to the one dimensional toy model , we have not the point like
singularity but the line like singurality
and the steepest descent line is also singular in this  case.
This makes a trouble in estimating the power of $N$ in $\alpha_k$.
But we show in the next section  numerical calculation indicate that
for even $k$ $\alpha_k$ has linear
scaling with respect to $N$.

For the case of odd $k$ , the saddle points can not completely ride on the
line of maximal singularity. But if we weaken the singularity, there is
a set of configurations:
\begin{eqnarray}
& &(m_1,m_2,m_3,m_4,\ldots ,m_{k-1} ,m_k ) \nonumber \\
&=& \{ (m,m,-m,m,\ldots ,m,-m),(m,-m,-m,m,\ldots ,m,-m),\nonumber \\
& & \quad (m,-m,m,m,\ldots ,m,-m),\ldots ,(m,-m,m,-m,\ldots ,m,m) \}
\end{eqnarray}
which we call ``next neutral''.
Under these configurations
$M_{i i+1}^{-1}=0 \quad (i=1,..,\check{j},..,k)$ are satisfied
by the following combinations of saddle points in the region of
 $A> m^2 \pi^2$
\begin{equation}
(t_1,\ldots,\check{t_j},\ldots ,t_k )=
(t_{\pm},t_{\mp},\ldots ,t_{\mp}) .
\end{equation}
Apparently the number
of the ``next neutral'' configurations is $k$  for fixed $m$.
These configurations
are dominant  in the large $N$ limit of $\alpha_k$.
The numerical calculation in the next section showes that
 $\alpha_{{\rm odd }}$ seems to oscilate with  $N^{-1/2}$ scaling .

Next we consider $\beta_k , \gamma _k$ i.e. the connected $k$-instanton
contribution to the Wilson loop.
Because we now consider large $N$ limit and the fact that the additive
imaginary charge in our representation of the Wilson loop has
order $1/N$ , the singular saddle point problem occure in the same manner as
$\alpha_k$. Thus we conclude that the dominant instanton configurations
 are  ``neutral''for $\beta_{{\rm even}},\gamma_{{\rm even}}$ and
``next neutral''for $\beta_{{\rm odd}},\gamma_{{\rm odd}}$.
The numerical calculation in the next section show that
$\beta_{{\rm even}},\gamma_{{\rm even}}$ have also linear scaling ($N$)
and $\beta_{{\rm odd}},\gamma_{{\rm odd}}$ have the oscilated $N^{-1/2}$
 scaling.

{}From the above argument , we find a peculiar phenomena about
the multi-instanton
configuration in the large $N$ limit.
We call the phenomena ``large $N$ neutrality''.
By ``neutrality'', we would not intend general neutral configuration which
satisfy $\sum_{i=1}^N m_i =0$, but intend that all instantons have equal
 absolute charge and number of positive charge instantons and number of
negative charge instantons are same or at most differs by one.

\section{Numerical calculation }

In this section we show the results of numerical calculation for the connected
multi-instanton amplitudes  ( $\alpha 's ,\beta 's ,\gamma 's $)
, the free energy
 and the Wilson loop for finite $N$.
We assume $0<A<4\pi^2$. From the above argument, in this region  we can
restrict the instanton charge as $m_{i}=\pm 1$
 in the multi-instanton sector.
In the numerical calculation we dont perform the  integral numerically
because we must deal
with highly fluctuated integral like the Fourier transformation.
But we  perform the residue summation which is finite series for finite $N$
 in our case.

First, we consider $\alpha_k $ .
Fig1 shows the numerical results of $\alpha_i (A) (i=1,\ldots ,4)$
with various $N$ and $A$.
We used the  following approximation .
For $\alpha_{{\rm even}}$  we restrict the ``neutral''
configuration eq.(56) with $m=\pm 1$ and
for $\alpha_{{\rm odd}} $    restrict the ``next neutral'' configurations
eq.(60) with $m=\pm 1$
which
are
dominant in the large $N$ limit.
\vskip10pt
\centerline{\epsfysize=2in\epsffile{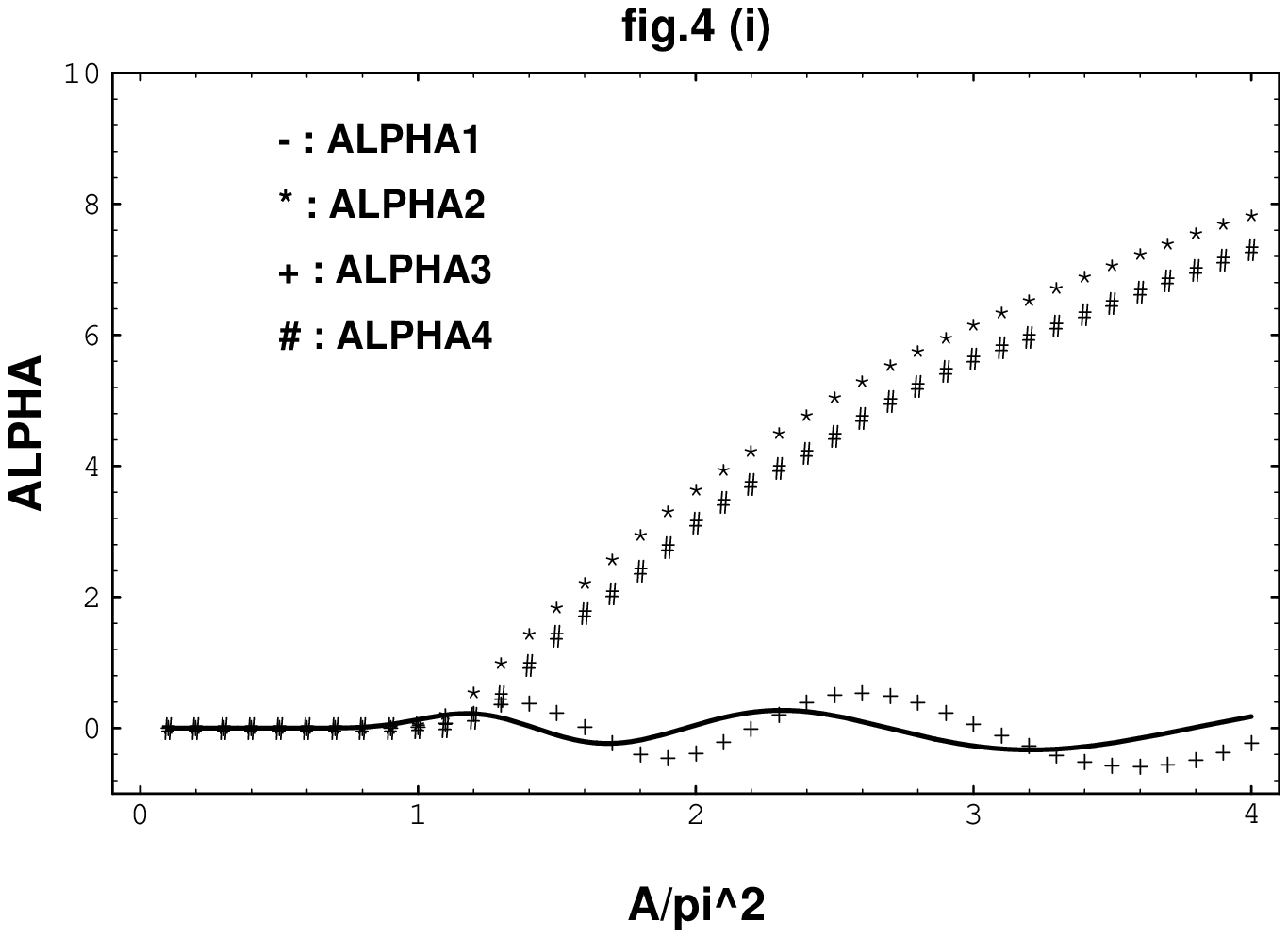} \hskip2pt
            \epsfysize=2in\epsffile{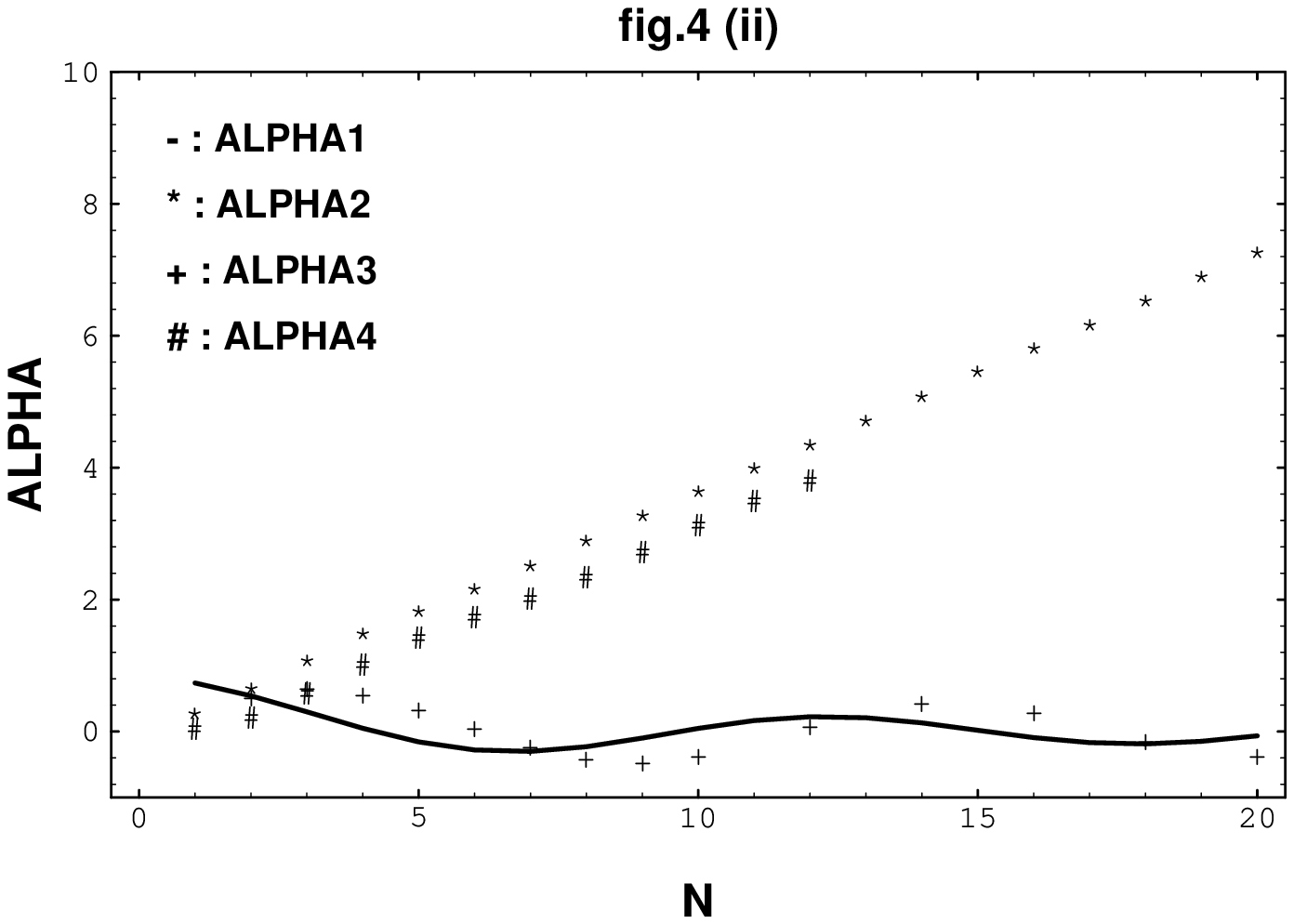}}
\vskip4pt
\hangindent\parindent{{\small {\bf Figure4}:
The connected multi-instanton amplitude $\alpha_j$ up to $j=4$
with (i) $N=10$ and various $A/\pi^2 $
and (ii) $A/\pi^2 =2$ and various $N$. There are remarkable difference between
 $\alpha_{{\rm even}}$ and  $\alpha_{{\rm odd}}$   . }}
\vskip10pt
We observe that compared to the $\alpha$ 's in $A>\pi^2$ , the $\alpha$ 's
in $A<\pi^2$ are negligible.
Moreover we see remarkable difference between
$\alpha_{{\rm even}}$ and $\alpha_{{\rm odd}}$
with respect to  order and shape in the region of $A>\pi^2$.
The shapes of $\alpha_2,\alpha_4$ are similar and both have the
properties of non-oscilating and linear scaling with respect to $N$.
On the other hand  $\alpha_1,\alpha_3 $
 have the properties of oscilating and $N^{-1/2}$ scaling .
We already observed that  $\alpha_1$
has $N^{-1/2}$ scaling and $ \alpha_2 $ has linear($N$) scaling using
 the large $N$ saddle point method \cite{o}.
Hence  from the argument in the previous section
 and these numerical results ,
we conjecture that  $\alpha_{{\rm even}}$ have linear ($N$) scaling but
$\alpha_{{\rm odd}}$ have $N^{-1/2}$ scaling in the large $N$ limit.
So there is ``democlacy'' in the interaction between the instantons. That is,
$\alpha_2$ which is the 2-body force , $\alpha_4$ which is the 4-body force
,$\ldots$ are same order with respect to $N$.

Figure5 shows the free energy  by our "neutral" multi-instanton scheme
\begin{equation}
\tilde{F}(A)=\frac{1}{N^2} \log [\frac{Z(A)}{Z_{{\rm weak}}(A)}] \to
\frac{1}{N^2} \log [ \oint \frac{ dz }{2\pi i}
     \frac{1}{z^{N+1}}
     \frac{1}{1-z} \exp(-\sum_{n=1}^{[N/2]} \frac{1}{2n} z^{2n}
     \alpha_{2n}(A) ) ]
\end{equation}
for $N$=3,4 and 5
 and the large $N$ exact solution of the free energy
obtained by Douglas-Kazakov. Here we neglect $\alpha_{{\rm odd}}$ from the
above observation.
\vskip10pt
\begin{center}
\centerline{\epsfysize=2.5in\epsffile{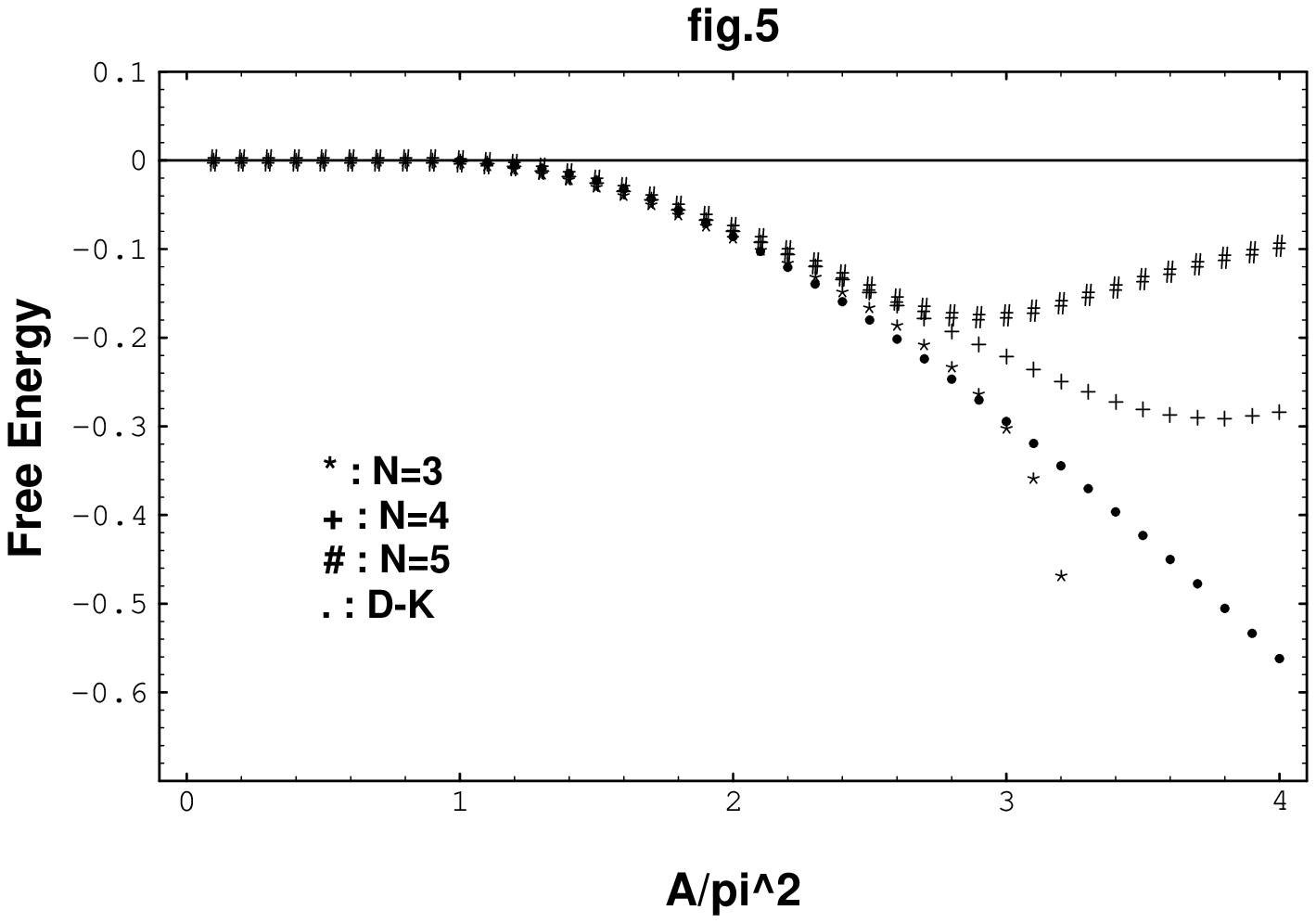}}
\vskip4pt
\begin{tabular}{|c|c|c|c|c|c|} \hline
   & $A=1.2\pi^2$ & $A=1.4\pi^2$ & $A=1.6\pi^2$ & $A=1.8\pi^2$ & $A=2\pi^2$
                                               \\ \hline
N=3&  0.009489    &-0.02186  &  -0.03949    &-0.06146 &   -0.08697
                                               \\ \hline
N=4&  -0.006966   &-0.01797  &  -0.03448    &-0.05545  &  -0.07973
                                               \\ \hline
N=5&  -0.005588   &-0.01580  &  -0.03173    &-0.05231  &  -0.07632
                                               \\ \hline
DK &  -0.005064   &-0.01497  &  -0.03213    &-0.05609  &  -0.08588
                                               \\  \hline
\end{tabular}
\end{center}
\vskip4pt
\hangindent\parindent{{\small {\bf Figure5}:
The free energy for $N=3,4,5$ approximated by the ``neutral''
configurations with  $m=\pm 1$
and the large $N$ exact solution by
Douglas-Kazakov.
The neutral multi-instanton picture well describe the behavior of the large
$N$ D-K solution in the neighborhood of the phase transition point.}}
\vskip10pt
The D-K solution is given by the following implicit equations:
\begin{eqnarray}
F'(A) &=& -\frac{a^2}{2}+\frac{1}{12}a^2 (1-\frac{b^2}{a^2})+\frac{1}{24}
          +\frac{1}{96}a^4 (1-\frac{b^2}{a^2})^2 A  ,\\
A     &=& 4 K(\frac{b}{a})
          (2 E(\frac{b}{a})-(1-\frac{b^2}{a^2}) K(\frac{b}{a}) ),\\
a     &=& 4 K(\frac{b}{a})/A
\end{eqnarray}
where $F(A)\equiv \frac{1}{N^2} \log Z(A) $ with the boundary condition
 $F(\pi^2)=\frac{1}{N^2} \log Z_{{\rm weak }}(\pi^2) $
and
$K(x),E(x)$ are the complete elliptic integral of first,
second kind respectively .
The above equations are obtained BIPZ-like analysis \cite{bipz}
 of the heat kernel representation  of the partition function.

The partiton function for $N$=3 in our approximation
turns into unphysical region in
$A{> \atop \simeq} 3.2 \pi^2$ and thus abort in this region.
We observe that the graphs for finite $N$ stand
up at $A\simeq \pi^2$ and in the neighborhood of  $A= \pi^2$ the
results for finite $N$ seem to converge on  the large $N$ D-K solution
as $N$ becomes large.
 But in the region $A{> \atop \simeq} 2\pi^2$ discrepancy of the finite $N$
 graphs and the  D-K solution becomes large.
Since we can see that
the further approximation with respect to absolute value of
 instanton charge does not make the discrepancy
small , we conjecture that the instanton summation  form of the partition
function is a kind of asymptotic series in the region of $A >\pi^2$.

Next we consider the $\beta_k ,\gamma_k$.
We follow the same approximation scheme as the $\alpha_k$. That is ,
 for $\beta_{{\rm even}},\gamma_{{\rm even}}$
we restrict the ``neutral''
configurations with $m=\pm 1$ and
for $\beta_{{\rm odd}},\gamma_{{\rm odd}}$
 restrict the ``next  neutral'' configurations with $m=\pm 1$.
\newpage
\vskip10pt
\centerline{\epsfysize=2in\epsffile{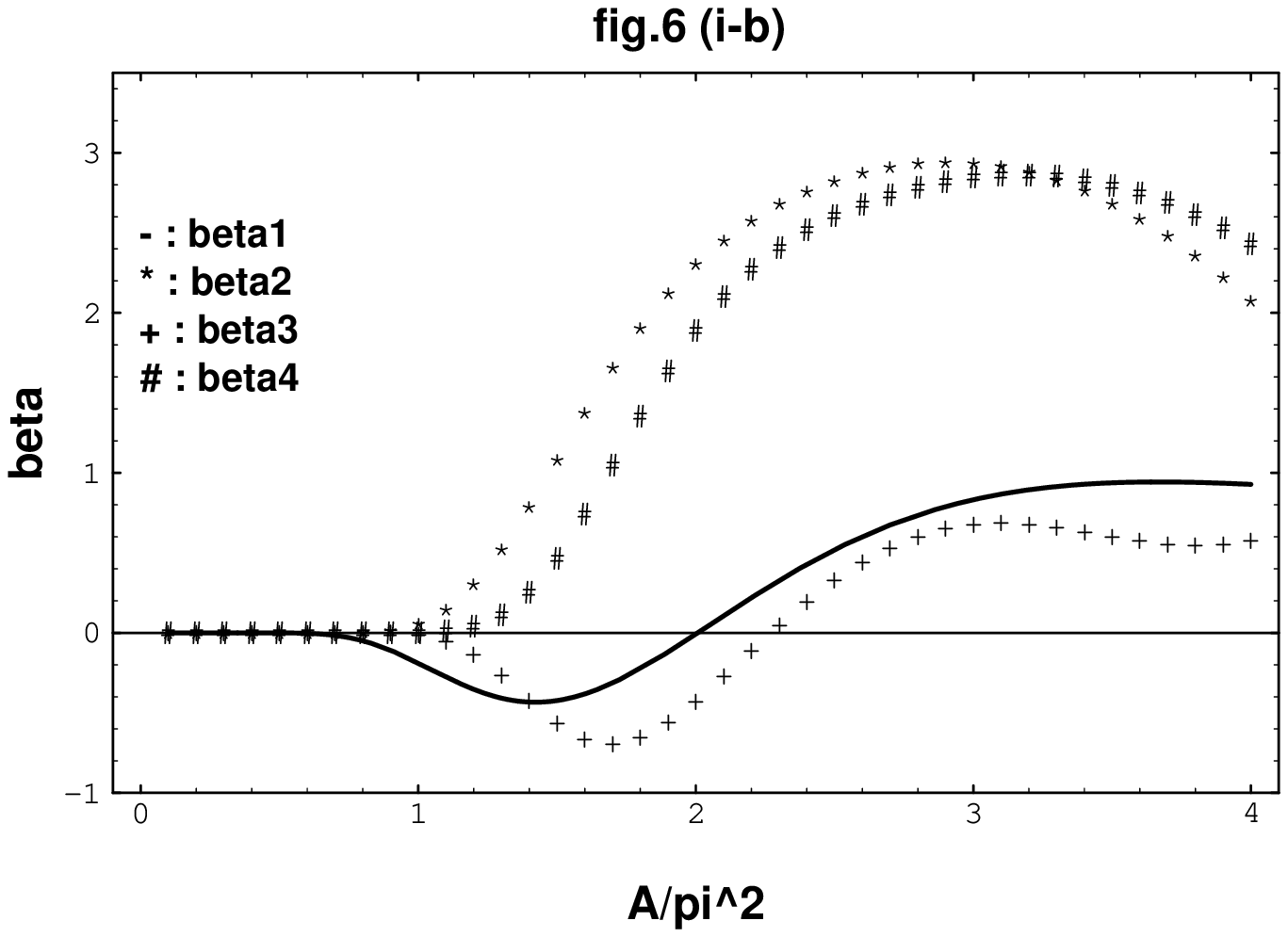}\hskip2pt
            \epsfysize=2in\epsffile{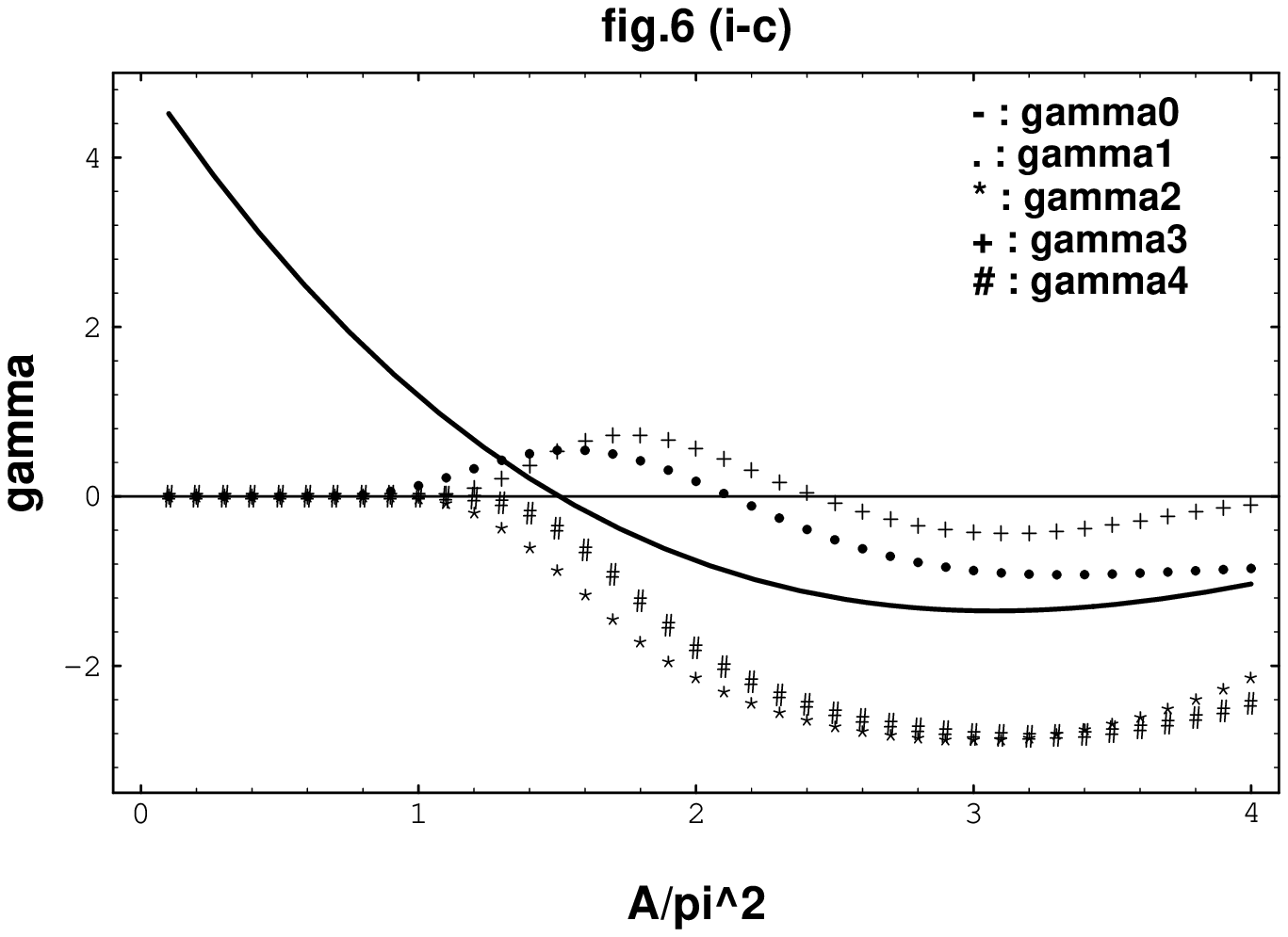}}
\vskip4pt
\centerline{\epsfysize=2in\epsffile{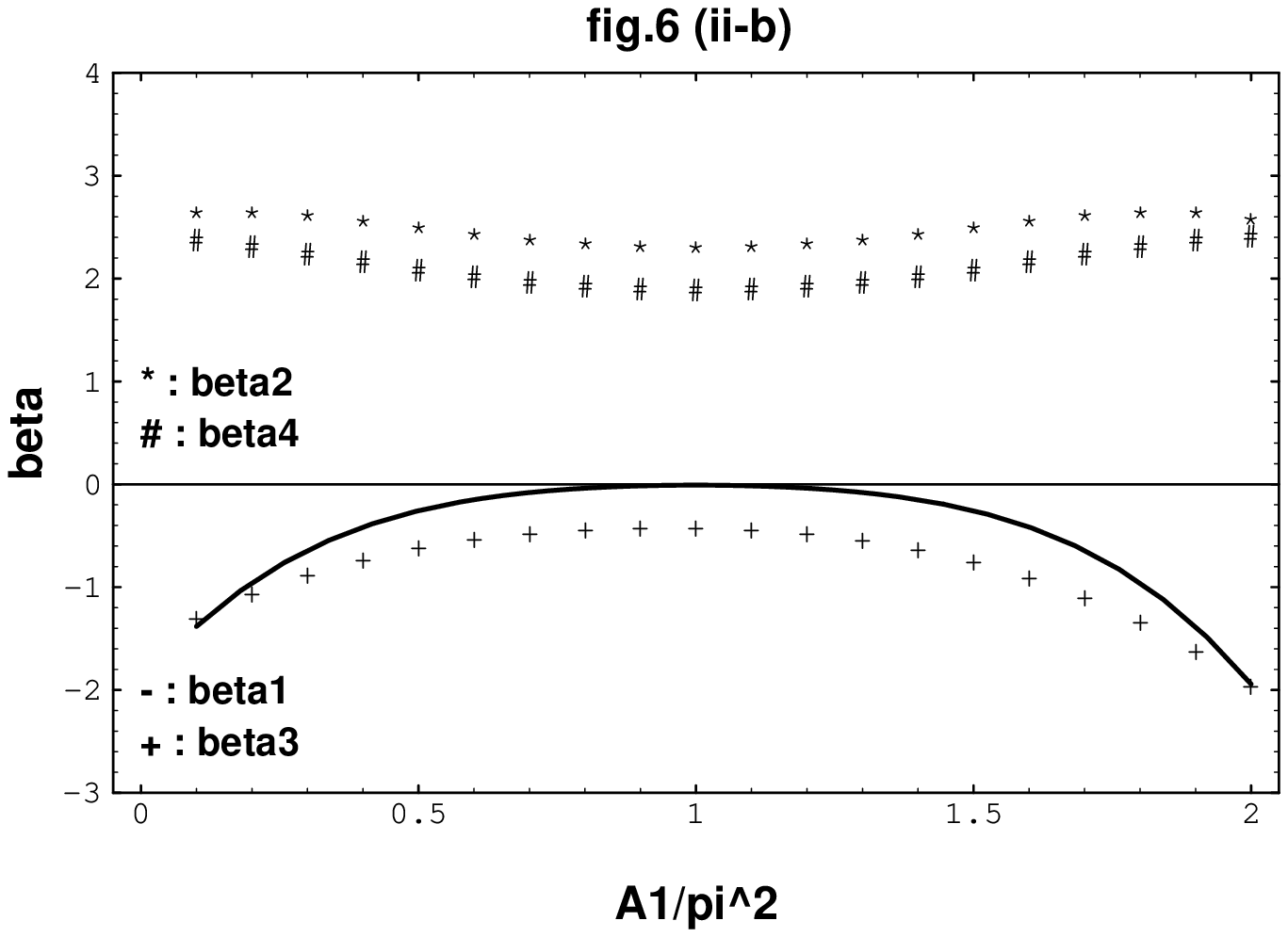}\hskip2pt
            \epsfysize=2in\epsffile{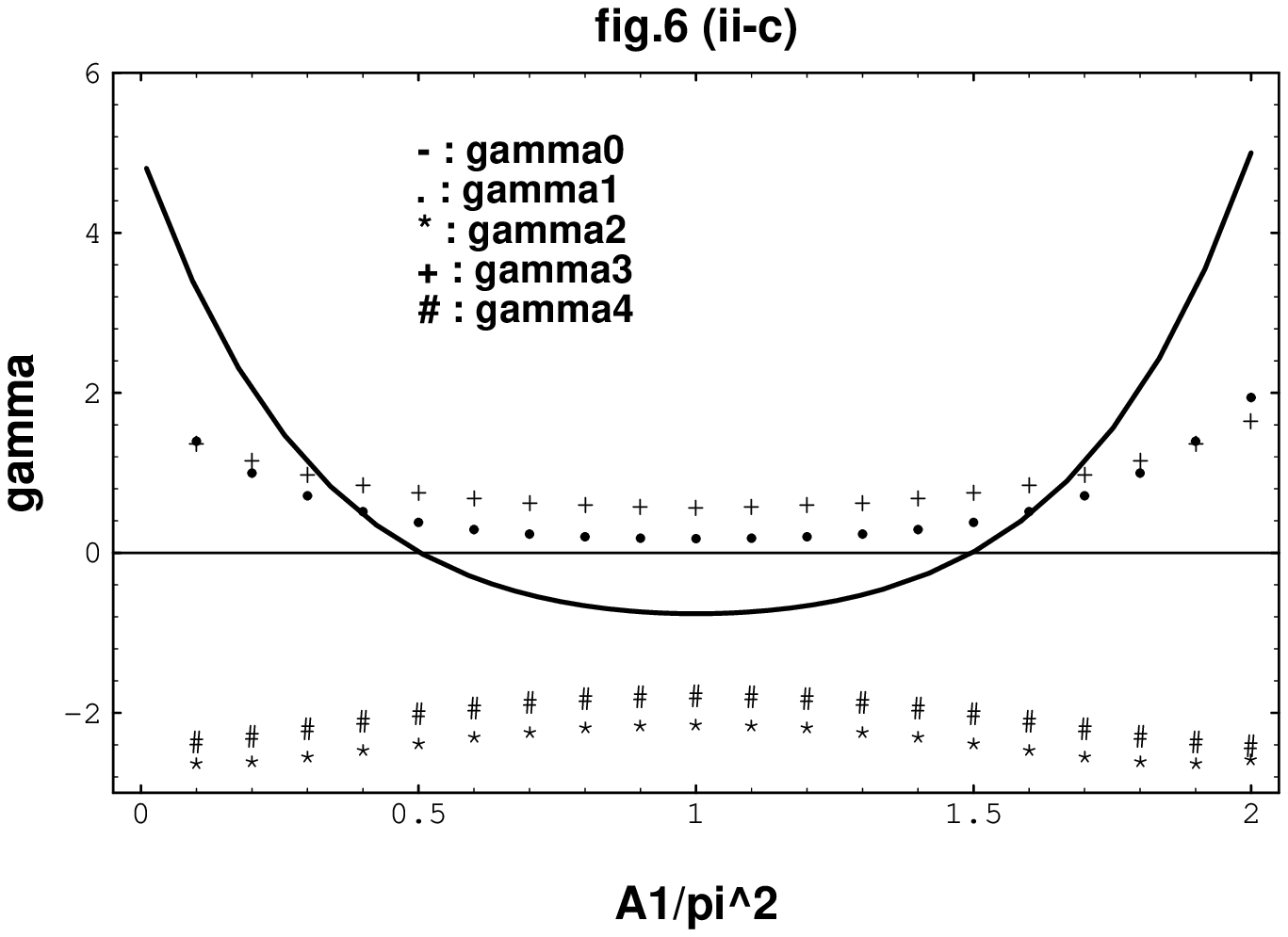}}
\vskip4pt
\centerline{\epsfysize=2in\epsffile{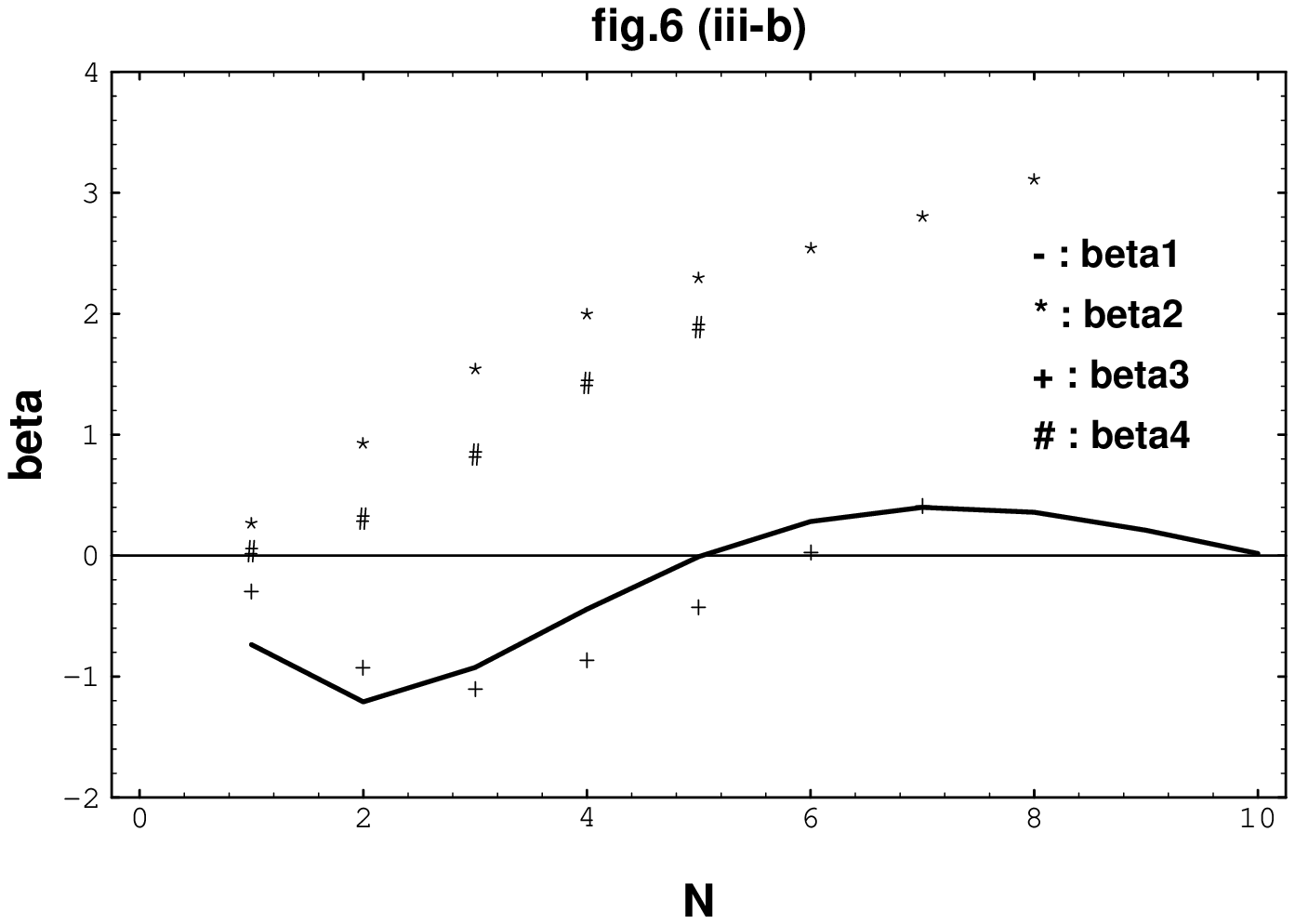}\hskip2pt
            \epsfysize=2in\epsffile{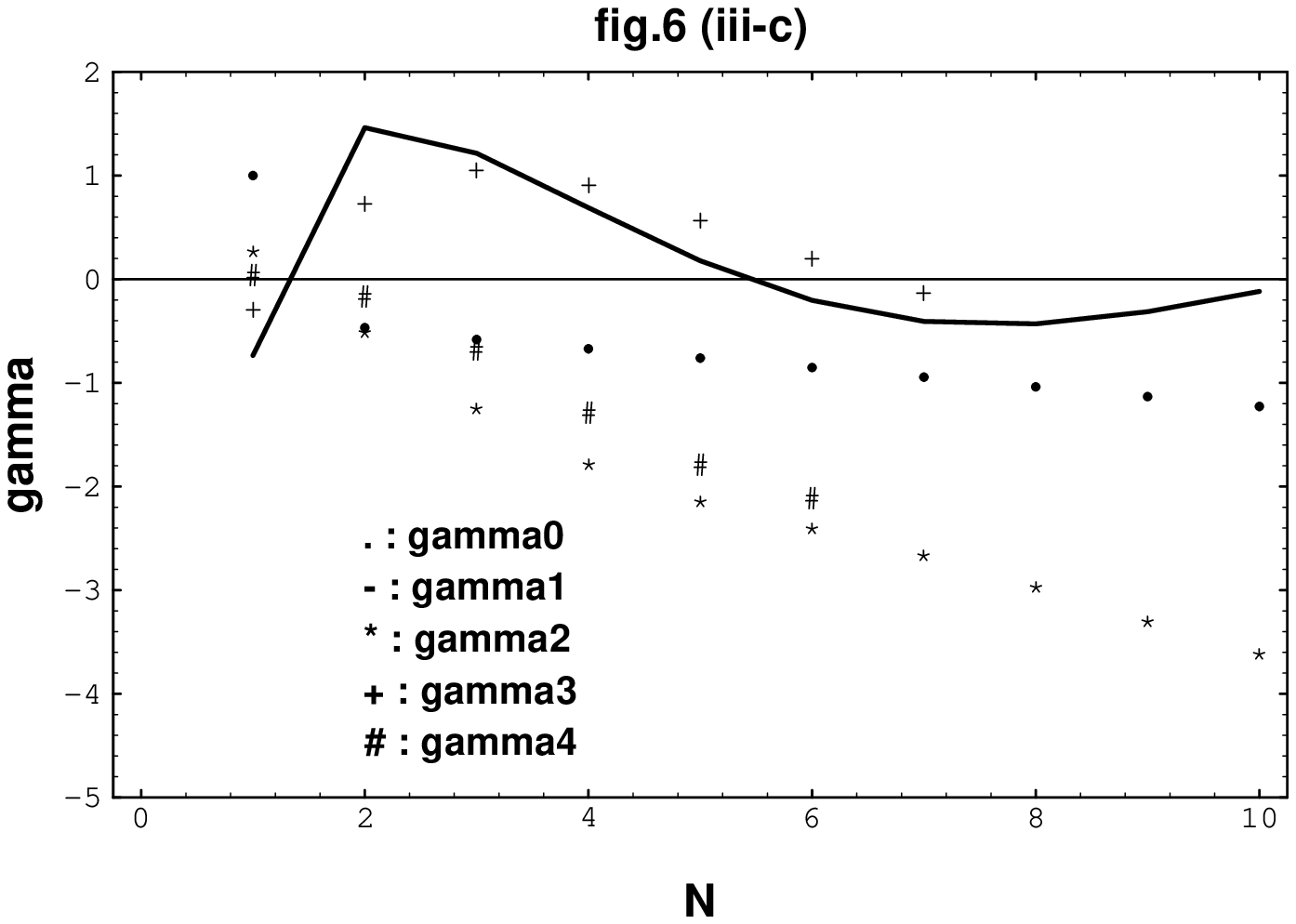}}
\vskip4pt
\hangindent\parindent{{\small {\bf Figure6}:
The contribution of the  connected multi-instanton amplitude
$\beta_j ,\gamma_j$to the
Wilson loop  with
(i) $N=5$ , various $A=2 A_1$
(ii)$ N=5 , A=2\pi^2$  various $A_1$
(iii)$A=2\pi^2 ,A_1 =\pi^2$ ,various $N$
 are shown }}
\vskip10pt
\newpage

Figure6 show $\beta_j  (j=1,\ldots,4)$ and
$\gamma_j  (j=0,\ldots,4)$ with
various $N,A,A_1$.
In fig.6  we observe that compared to the $\beta_j ,\gamma_j $
in $A>\pi^2$ ,
 the $\beta_j ,\gamma_j $ in $A<\pi^2$ are also negligible.
But compared to $\alpha_j$  there is no remarkable difference
with respect to order
between  $\beta_{{\rm even}}$ and $\beta_{{\rm odd}}$ and between
$\gamma_{{\rm even}}$ and $\gamma_{{\rm odd}}$ in figure6(i)(ii).
We should note these are $N=5$ results.
And we show in  fig.6(iii) both $\beta_{{\rm even}}, \gamma_{{\rm even}}$
 have  linear
scaling with respect to $N$ contrasted to oscilated damping behavior of
$\beta_{{\rm odd}}, \gamma_{{\rm odd}}$ .
Thus in the large $N$ limit  only  $\beta_{{\rm even}},\gamma_{{\rm even}}$
 are
survived and the approximation in which
$\beta_{{\rm odd}}, \gamma_{{\rm odd}}$ are neglected
 fot finite $N$ is more
nearer to the large $N$
exact solution than the approximation  in which
all  $\beta,\gamma$ are included.

Figure7 show the  simple Wilson loop average (i.e. winding  $n=1$ )
for $N=3,4,5$ by our ``neutral'' multi-instanton scheme :
\begin{eqnarray}
& & \widetilde{W}_{1}(A_1,A_2)   \\
&=& \ee^{-\frac{ A_1 A_2}{2AN}} [
    \frac{  \frac{1}{N} \sum_{l=0}^N \oint \frac{dz}{2\pi i}
     ( \sum_{k=0}^{l-1}\frac{\beta_{l-k}(A_1,A_2)}{z^{k+1}}+
       \sum_{k=0}^{l-1} \frac{\gamma_{l-k}(A_1,A_2)}{z^{k+1}} )
      \exp (-\sum_{n=1}^{[\frac{N}{2}]}
             \frac{1}{2n} z^{2n} \alpha_{2n}(A) )}
     {\oint \frac{dz}{2\pi i} \frac{1}{z^{N+1} (1-z)}
      \exp (-\sum_{n=1}^{[\frac{N}{2}]}
             \frac{1}{2n} z^{2n} \alpha_{2n}(A) )}]
        \nonumber
\end{eqnarray}
where  $\beta_{{\rm odd}},\gamma_{{\rm odd}}$ are neglected
and  the large
 $N$ exact solution obtained by
Boulatov and Daul-Kazakov (BDK)\cite{bdk}.
\vskip10pt
\begin{center}
\centerline{\epsfysize=2.5in\epsffile{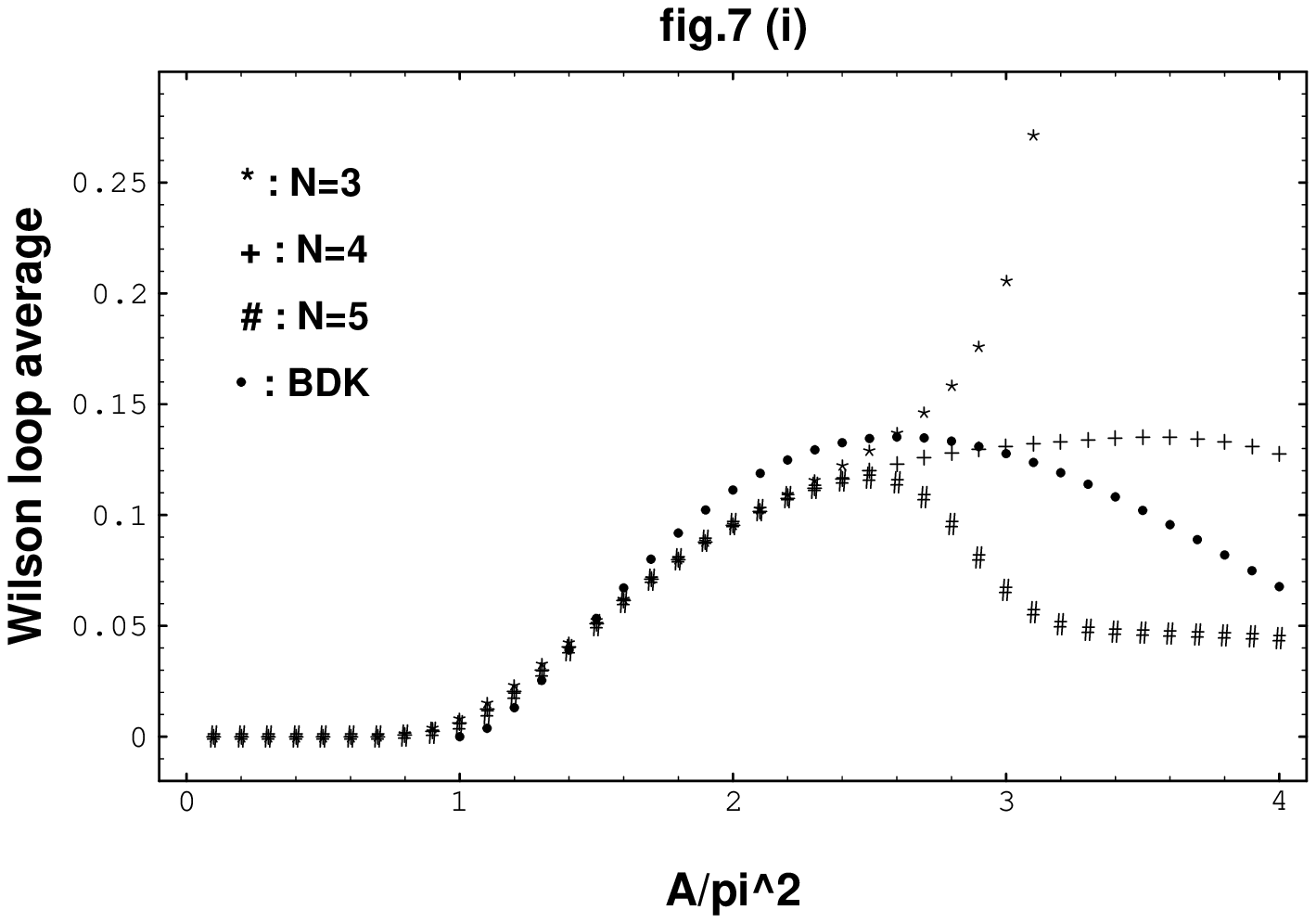}}
\begin{tabular}{|c|c|c|c|c|c|} \hline
   & $A=1.2\pi^2$ & $A=1.4\pi^2$ & $A=1.6\pi^2$ & $A=1.8\pi^2$ & $A=2\pi^2$
                                               \\ \hline
N=3& 0.02343 & 0.04265  & 0.06227     &0.08011 &  0.09564
                                               \\ \hline
N=4& 0.02049 & 0.04051  &  0.06132    & 0.07992 & 0.09525
                                               \\ \hline
N=5& 0.01861 & 0.03919  &  0.06085    & 0.08014 &   0.09589
                                               \\ \hline
BDK& 0.01307 & 0.03918  &  0.06712    &0.09186  &  0.11130
                                               \\  \hline
\end{tabular}
\centerline{ \epsfysize=2.5in\epsffile{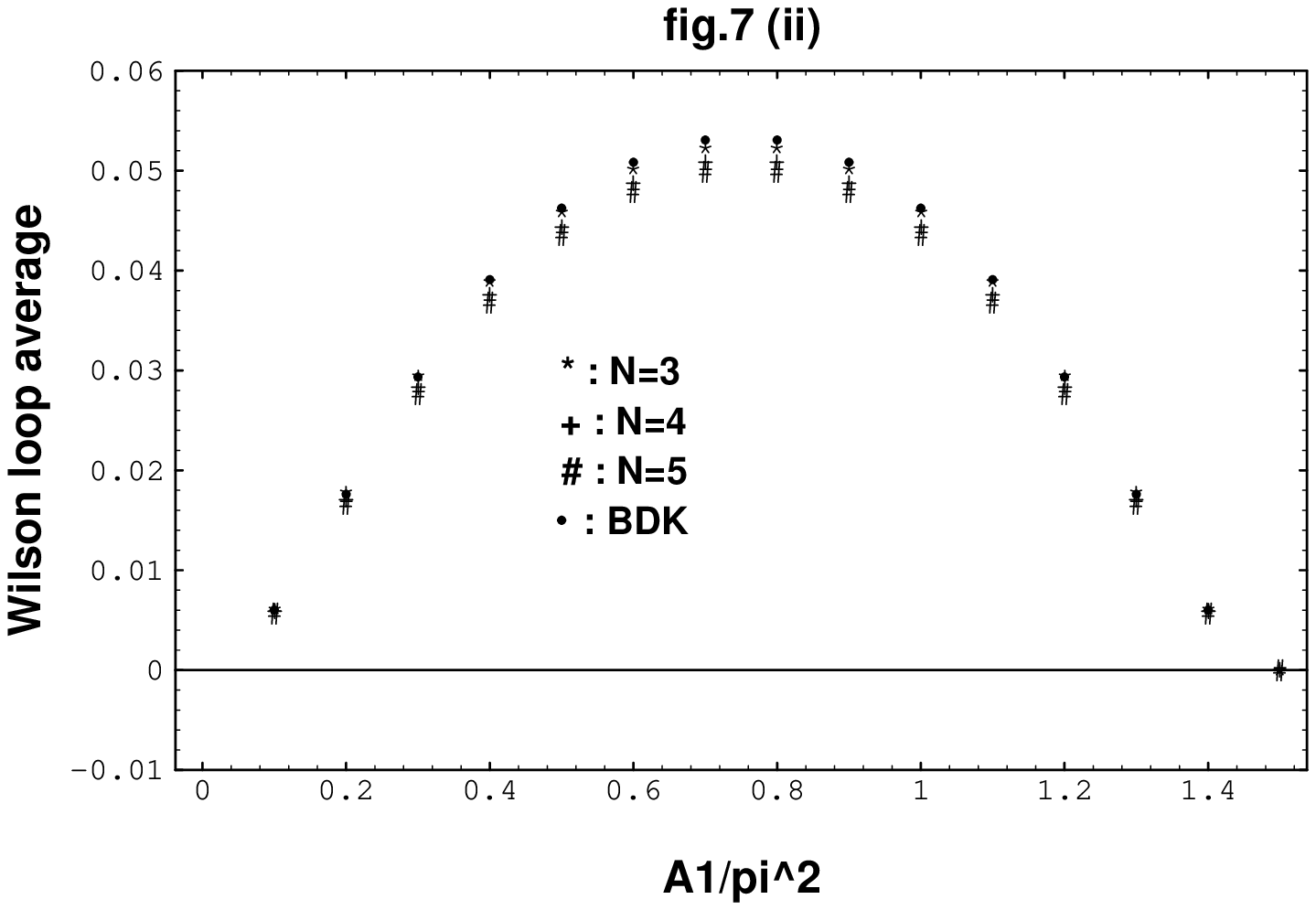}}
\end{center}
\vskip4pt
\hangindent\parindent{{\small {\bf Figure7}:
The Wilson loop average with
(i) $N=3,4,5$ , various $A=2 A_1$
(ii)$ N=3,4,5 , A=1.5\pi^2$  various $A_1$
  by our ``neutral'' multi-instanton scheme
and for each case showed the large $N$ Boulatov,Daul-Kazakov solution.
The neutral multi-instanton scheme well describes the behavior of the large
$N$ BDK solution in the neighborhood of the phase transition point} }
\vskip10pt
The large $N$ BDK solution is also given by
the following implicit equation :
\begin{equation}
W_1 (A_1,A_2)=\oint_C  \frac{dz}{2\pi i}\ee^{ A_1 z - f(z)} .
\end{equation}
where $f(z)$ is the  resolvent  in this system and the contour $C$
encircles the cut of $f(z)$ countercloskwise. From the above
equation the difference of the phase in the Wilson loop average
is concentrated on the resolvent . The resolvent is given by the
 famous Wigner's semi-circle law
 in the weak coupling phase ($A<\pi^2$):
\begin{equation}
f(z)\equiv \frac{Az}{2} -\frac{A}{2}\sqrt{z^2-4/A}
\end{equation}
On the other hand in the strong coupling phase the resolvent is given by
\begin{equation}
f(z)\equiv \frac{Az}{2} +\frac{2}{az} \sqrt{(a^2 -z^2)(b^2-z^2)}
           \pi (-\frac{b^2}{z^2},\frac{b}{a})
\end{equation}
where $a,b$ are the same parameter as the
equations of the free energy and $\pi (c,x)$ is the complete elliptic integral
of third kind.

We observe that the graphs for finite $N$ stand
up at $A\simeq \pi^2$ and in the neighborhood of  $A= \pi^2$ the
results for finite $N$ seem to converge on  the large $N$ BDK solution
as $N$ becomes large.
But  we also observe that the disagreement becomes large
 in the region $A> 2 \pi^2 $ which is
same as the free energy. Hence the instanton summation form of the Wilson
loop  is also a kind of asymptotic series and divergent in the region of
$A> \pi^2 $.

Hence  we  conclude that viewed from weak coupling phase
 the large $N$
phase transition is physically caused by the ``neutral''
 multi-instanton.

\section{conclusion}

We have analyzed the multi-instanton amplitudes in terms of
 the contour integral representation and found that the represenation
make clear
the properties and the role of the multi-instanton in the large $N$ phase
transition.
In particular the ``neutral'' configurations of the even number instantons
 are essential.
We also showed that our ``neutral'' multi-instanton scheme well describes the
behavior of the free energy and the Wilson loop around the critical point
$A_c =\pi^2$ up to $A \simeq 2\pi^2$.
And we show that is imposibble in the dilute gas approximation.

But we observe the agreement region of our results and the large $N$ exact
solution tends to become narrow as $N$ becomes large.
In addition we can see that the further approximation with respect to
absolute value of instanton charge does not make the discrepancy
small.
So we conjecture the form of the instanton summation is a kind of asymptotic
series.
For example ,assume that $S(z)$ is defined nonperturbatively and has an
asymptotic expansion $S(z) \Rightarrow \sum_{k=0}^{\infty} s_k z^k$.
By truncating the series with apropriate region of
summation in the neighborhood of $z=0$ ,
we can approach to the true value.
The same scenario works for our case of the ``neutral'' multi-instanton
truncation.
Thus we conclude that our  ``neutral'' multi-instanton scheme
is a good
truncation in the strong coupling region  up to $A \simeq 2\pi^2$
 at large $N$.

\vskip10pt
Recently the field strength correlators in the large $N$ limit are determined
\cite{ns}
using the abelianization technique for the path integral\cite{bt}.
To explore whether our method can describe the behavior of these correlators
is our next task.

\newpage
\begin{center}
{\large {\bf Acknowledgements}}
\end{center}
\vskip10pt
The auther would like to thank Professor K.Ishikawa for valuable discussions.

\end{document}